\documentclass[twocolumn]{aa}
\usepackage{graphicx}
\usepackage{natbib}
\usepackage{verbatim}
\usepackage{txfonts}
\begin{document}
\newcommand{\gcu}{\mbox{$\, \stackrel{ > }{ _{\sim} } \,$}}
\newcommand{\lcu}{\mbox{$\, \stackrel{ < }{ _{\sim} } \,$}}
\title{Uncertainties on the theoretical predictions for classical Cepheid pulsational quantities}

\titlerunning{Uncertainties on classical Cepheid predictions}

\author{G. Valle \inst{1}, M. Marconi\inst{3}, S. Degl'Innocenti\inst{1,2}, P.G. Prada Moroni\inst{1,2}
}

\authorrunning{Valle G. et al.}

\institute{
(1) Dipartimento di Fisica ``Enrico Fermi'', 
Universit\`a di Pisa, largo Pontecorvo 3, Pisa I-56127 Italy
\and
  INFN,
 Sezione di Pisa, Largo B. Pontecorvo 3, I-56127, Italy
 \and
INAF, Osservatorio Astronomico di Capodimonte, via Moiariello 16, 80131
  Napoli, Italy}

\offprints{M. Marconi, marcella@na.astro.it}
      
\date{28/09/2009}

 \abstract 
{With their period-luminosity relation, ``Classical Cepheids'' (CC) are the most common primary distance
     indicators within the Local Group, also providing an absolute calibration
     of important secondary distance indicators. However, the predicted position of these pulsators in the HR diagram, along the so called {\it blue loop}, that is the expected distribution of Cepheids within the
     instability strip is affected by several model inputs, reflecting upon the predicted $PL$ relation.}  
   {The aim of this work is to quantitatively evaluate the effects on the
     theoretical $PL$ relation of current uncertainties on the chemical
     abundances of Cepheids in the Large Magellanic Cloud (LMC) and on several physical assumptions
     adopted in the evolutionary models.  We will separately
     analyse how the different factors influence the evolutionary and
     pulsational observables and the resulting $PL$ relation.}
  {To achieve this goal we computed new sets of updated evolutionary and pulsational models.}
   {As a result, we find that present uncertainties on the most relevant H and He
     burning reaction rates do not influence in a relevant way the loop
     extension in temperature. On the contrary, current uncertainties on the LMC chemical composition significantly affect the loop extension and also reflect in the morphology of the instability strip; however their influence on the predicted pulsational parameters is
 negligible. We also discussed how overshooting and mass loss, sometimes
 suggested as possible solutions for the long-standing problem of the Cepheid
 mass discrepancy,  influence the $ML$ relation and the pulsational parameters.} 
  {In summary, the present uncertainties on the physical inputs adopted in the
 evolutionary codes and in the LMC chemical composition are negligible for the
    prediction of the main pulsational properties. On the other hand, the inclusion of overshooting in the previous hydrogen burning phase and/or of mass loss is expected to significantly change the resulting theoretical pulsational scenario for Cepheids, as well as the calibration of their distance scale. These systematic effects are expected to influence the theoretical Cepheid calibration of the secondary distance indicators and in turn the resulting evaluation of the Hubble constant.}

   \keywords{cepheids -- stars:evolution -- stars:distances} 
   
   \maketitle

\section{Introduction}\label{sec:intro}

During the central helium burning phase, intermediate mass stars show an
excursion toward higher effective temperature, in the Hertzprung-Russell (HR)
diagram, subsequently coming back toward the asymptotic giant branch ({\em blue
  loop}). During this phase the stars cross the instability strip, becoming
Classical Cepheids.  These variable stars, thanks to their characteristic
Period-Luminosity ($PL$) relation, are the most common primary distance
indicators within the Local Group. Moreover, from space, they are
observable also in external galaxies, up to distances of 20-30 Mpc \citep[see
e.g.][]{free01}. On this basis they provide the absolute calibration of
important secondary distance indicators, such as the maximum luminosity of
Supernovae Ia, the Tully-Fisher relation, surface brightness fluctuations,
and the planetary nebulae luminosity function, that are the basis for measurement
of the Hubble constant \citep{free01,sa01}.  Any
systematic error affecting the Cepheid $PL$ relations affects the
extragalactic distance scale and the estimate of the Hubble
constant. Therefore, even though the Cepheid $PL$ relation is 
assumed to be universal and calibrated through the sample in the Large
Magellanic Cloud \citep[LMC, see e.g.][]{mf91,u99}, the possibility that the
Cepheid properties and $PL$ relation depend on the chemical composition of the
parent galaxies has been actively debated during the last decade \citep[see
e.g.][]{S04,M06,r05,r08,M05,b08}. On the other hand the theoretical
prediction for the existence of a $PL$
relation for Classical Cepheids relies on the assumption that intermediate
mass stars undergoing central helium burning are characterized by a
mass-luminosity ($ML$) relation, as predicted by stellar evolution models. 
On very general grounds the pulsation period is expected to be tightly
correlated with the luminosity, the mass and the effective temperature of the
star, for each given chemical composition. It is only thanks to the $ML$
relation that the period can be simply related to the luminosity and the
effective temperature, allowing a theoretical prediction for the period-luminosity-color relation
\citep[$PLC$, see][]{f84,f91,f95,cc86,ls86,s88,cl91,s97}.  However the
theoretical $ML$
relation depends on the several physical processes included in stellar
evolutionary codes as well as on the adopted chemical composition.

By averaging the $PLC$ relation on the color extension of the pulsation
instability strip one obtains the $PL$ relation. This implies that the $PL$
relation has a statistical nature and observationally holds only when a
statistically significant sample of Cepheids (at the same distance) is
available. Thus the predicted temperature
extension of the blue loop for a given mass also reflects upon the expected
distribution of Cepheids within the instability strip and in turn upon the
theoretical $PL$ relation.  This is a very critical point because the
predicted extension
of the blue loop in the HR diagram is affected by several model inputs
(chemical composition, efficiency of external convection, nuclear burning
cross sections, etc..) producing effects that are not yet completely understood
\citep[see e.g.][]{Xu2004a,stot94,bru90}. In this context, a detailed
comparison of evolutionary and pulsational models with the observations can
provide insight into the physics of stellar evolution and pulsation.

The aim of this work is to quantitatively evaluate the effects on the
predicted $PL$ relation of current uncertainties on the chemical
abundances of Cepheids in the LMC and of several physical assumptions
adopted in the evolutionary models.  In particular, we will separately
analyse how the different factors influence the evolutionary and
pulsational observables (extension in effective temperature of the
blue loop, Mass-Luminosity relation, morphology of the instability
strip, period distribution) and the resulting $PL$ relation. For
the chemical composition we choose that of the LMC with its associated
uncertainty.  All the relations based on the newly computed sets of
evolutionary and pulsation models will be made available to the
community.  In a forthcoming paper (Valle et al. 2009 in preparation) we
will focus on the comparison of these new predictions with the
observations.  The organization of the paper is the following: in
Sect.~\ref{sec:2} we present the theoretical scenario both from the evolutionary
and the pulsational point of view; Sect.~\ref{sec:blueloop} contains a general
discussion about the problems in calculating blue loop models; in
Sect~\ref{sec:rates} the dependence of the
predicted extension of the {\it blue loop} on the uncertainties
affecting nuclear reaction rates is investigated; in Sect.~\ref{sec:deltaz}
we focus on the effects of chemical abundance uncertainties on the most relevant evolutionary and pulsational features, whereas in Sect.~\ref{oversh} we
take into account possible noncanonical effects in the evolutionary scenario. The conclusions,
reported in Sect.~\ref{sec:concl}, close the paper.

\section{Theoretical scenario}\label{sec:2}

In this section we present the new sets of evolutionary and pulsation models
that have been used for the calculations.

\subsection{Evolutionary stellar models}

Evolutionary tracks were computed with an updated version of the FRANEC
evolutionary code \citep[see e.g.][]{chieffi,deg08} including
the OPAL 2006 equation of state (EOS) \footnote{http://www-phys.llnl.gov/Research/OPAL/EOS$\_2005/$}
\citep[see also][]{rog96} and  
radiative opacity
tables\footnote{http://www-phys.llnl.gov/Research/OPAL/opal.html}
\citep[see also][]{ig96} for temperatures higher than
12000 K. Moreover our models include the conductive  
opacities by \citet{shtern} \citep[see also][]{pot} and the
atmospheric opacities by \citet{ferg} 
\footnote{available at http://webs.wichita.edu/physics/opacity}.
All the adopted  opacity tables have been calculated by assuming the solar relative
metal abundances by \citet{aspl05}.

Since the $^{12}$C($\alpha$,$\gamma$)$^{16}$O and $^{14}$N(p,$\gamma$)$^{15}$O
reaction rates affect the evolutionary properties of intermediate-mass stars
and in particular the temperature blue loop extension \citep[see
e.g.][]{bru90}, we adopted the updated rates by \citet{12c} for the
$^{12}$C$+ \alpha$ and by the LUNA collaboration \citep[see][and references
therein]{14n} for the $^{14}$N$+$p cross sections.  For the other nuclear reactions we
adopted the same rates as in \citet{Ciacio97}.  The code implements weak screening
\citep{Salpeter54}, weak-intermediate, intermediate-strong screening
\citep{Graboske}, and strong screening \citep{Itoh1977,Itoh1979}.

For convective mixing, we adopt the Schwarzschild criterion to define regions
in which convection elements are accelerated 
\citep[see the description in][]{broc03} with an additional algorithm to take into
account the induced 
semiconvection during the central He-burning phase \citep{castellani71a,castellani71b}.
 Rotationally induced effects \citep[see e.g.][]{maed98,pala03} are not included in our models.

To model envelope convection we adopted, as usual, the mixing length formalism
\citep{bohm58}. The mixing length parameter, $\alpha$, governing the efficiency
of convection, has been calibrated to reproduce the observed stellar colors in
the Large Magellanic Cloud cluster NGC 1866, obtaining $\alpha$=1.9 \citep{broc04}. 
This result is also dependent on the atmospheric models adopted to transform
evolutionary calculations from the theoretical (logL-logT$_e$) to the
observational plane. In this paper we adopted the color-temperature transformations by \citet{bess}.

The original helium mass fraction ($Y$) cannot be measured
spectroscopically and it is generally estimated by assuming a linear relation between $Y$ and $Z$:
\begin{equation}\label{eq:elio}
 Y = Y_P + \frac{\Delta Y}{\Delta Z} \; \times \; Z \quad
\end{equation}
where $Y_P$ is the primordial helium abundance and $\Delta Y$/$\Delta Z$ is
the estimated helium-to-metal enrichment ratio. For $Y_P$ we adopted the value
$Y_P = 0.248$ \citep{Peimb,Izot} and $\Delta{Y}/ \Delta{Z} = 2$ \citep{dydz}.

The metallicity adopted for the models is consistent with the most recent estimates of [Fe/H] for LMC Cepheids \citep[see][]{luck98,r08} and also with several independent determinations of the LMC metallicity \citep[see e.g.][]{k00,a01}.

The total metallicity can be expressed as follows:

\begin{equation}
\label{eq:zeta}
 Z = (1-Y_P)\left(1+\frac{\Delta Y}{\Delta Z}+ \frac{1}{(Z/X)_\odot} \times 10^{-[Fe/H]}\right)^{-1} \quad 
\end{equation}
where $(Z/X)_\odot$ is the solar metal to hydrogen abundance ratio.

In this work we adopt the updated \citet{aspl05} mixture, based on the analysis of spectroscopic data by
means of tridimensional hydrodynamical atmospheric models, instead of that by
\citet{gre93}, hereinafter GN93, widely used in the past. This
leads to a significant reduction of the the $Z/X$ ratio for the Sun, from
$Z/X \approx 0.0245$ to $Z/X \approx 0.0165$. As a result the traditionally assumed metallicity
for LMC decreases from $Z \approx 0.008$ to $Z \approx 0.005$. With this choice of
metallicity, from Eq.~\ref{eq:elio}, one obtains $Y \approx 0.26$, which we
assume as our standard value.

Several investigations have dealt with the analysis of the effects of this
revision of the solar abundance, revealing discrepancies between predicted and
helioseismologically inferred quantities \citep[see e.g.][]{bah05,basu04}.
This situation is further complicated by the fact that current uncertainties
on the physical inputs adopted in the solar models can affect the results of
the quoted investigations. Moreover some authors questioned the accuracy of
the oxygen abundance derived by \citet{aspl05} \citep[see e.g.][]{caff08}.
For these reasons, even if we adopted the \citet{aspl05} solar mixture as our
reference mixture, in the following we will take into account the uncertainty
related to this assumption and its influence on the LMC metallicity.

\subsection{Nonlinear pulsation models}

Our pulsational models are based on a nonlinear nonlocal
time-dependent convective hydrodynamical code 
\citep[][and references therein]{BS94,BMS99}. This code has been successfully applied to several
classes of pulsating stars \citep[see e.g.][and references therein]{M03,DC04,B99,M04,M05,M07}. In this paper the physical ingredients of Cepheid pulsation models,
and in particular the stellar opacity tables, have been updated in order
to insure consistency between the 
 pulsation and the evolutionary models.  The new
pulsation models have been constructed with the same chemical
abundances of evolutionary models, covering the same range in stellar mass and
assuming the luminosity predicted by the canonical
Mass-Luminosity ($ML$) relation by \citet{B00}. For $Z=0.005$, a second
luminosity brighter by 0.25 dex has been included to account 
for the effect of mild overshooting and mass loss (see Subsec.~\ref{oversh}).

\section{Modeling the blue loop morphology}\label{sec:blueloop}
The extension in
temperature of the blue loop is an important evolutionary feature. Stellar models show that
the excursion to the blue is produced by an envelope contraction
\citep[see e.g.][]{renz92}: models with more extended 
loops undergo a more significant envelope contraction which heats the hydrogen
shell, increasing its luminosity. The return of the track to the red is due to 
a variety of effects \citep[see e.g.][]{sesti02} : a) when the central
helium abundance decreases below $Y_{c}\sim0.1$ the star reacts to the
decreased burning rate by means of a core contraction increasing again the 
energy
production and thus the amount of energy reaching the envelope, which 
expands and cools;  b) the H shell reaches the discontinuity in the
chemical profile left by the first dredge up before the core
contraction \citep[see e.g. the discussion in][]{lau71,hua83,sesti02,Xu2004a}.
However, as widely discussed in the literature \citep[see
e.g.][]{stot93,alc78,rob71,bru90,sesti02,Xu2004b}, the formation and
the extension of loops appear to be caused
by the interaction of several factors and is extremely sensitive to small 
changes 
either in the stellar chemical composition or in the physical inputs adopted in 
the calculations.  Several authors have discussed the poor stability naturally existing in many stellar models which lie on the red
supergiant branch during the central He burning phase, that is, around the
beginning of the blue loop \citep[see
e.g.][]{NoelGabr73, StoChi91a}.

The dependence of the loop morphology on changes in the chemical composition will be
analysed in Sec.~\ref{sec:deltaz}. In general, an
increase of the original He abundance increases the mean molecular weight
while it leads to a decrease of the mean opacity \citep[see e.g.][]{VemSto78},
even if the dependence on He abundance is not linear with metallicity.  On the other hand a 
metallicity variation changes the blue loop extension mainly due to the
opacity variation
\citep[see e.g.][]{stot93} even if the behaviour can be more complicated
and a quantitative explanation of the loop
dependence on chemical composition is still lacking in the literature.

The evolutionary properties of intermediate and high mass stars in the central
He burning phase strongly depend on the efficiency of the He burning nuclear
reactions \citep[see e.g.][]{Iben72, bru90, B00}. This point is
discussed in the following subsections in the light of updated nuclear cross
sections measurements.

Regarding the opacity dependence, any increase of the opacities deep in the
stellar interiors could lead to fainter luminosities, cooler effective
temperatures and longer lifetimes.  Regarding the loop extension, in
principle, a higher opacity leads to a reduced blue loop but the effect is in
general very small \citep[see e.g.][]{StoChi91b}. In addition, updated opacity
tables agree amongst each other within 5-10\% \citep[see e.g. discussion
in][]{Bad05, Neu01}
thus very small opacity uncertainties are expected.
Moreover, molecular opacities have almost negligible effects
on the blue loop extension  \citep[see e.g.][]{stot93}.

Several authors \citep[see e.g.][]{Chio92,StoChi92,BroCast93} noticed that the
extension of the convective H burning core affects the following He burning
phase. This involves the convective core overshooting efficiency, whose
influence on the blue loop extension will be discussed in detail in Sec.~\ref{oversh}
and the method adopted for treating the semiconvective layers that develop
around the H burning core close to the central H exhaustion. 
If a significant amount of overshooting is included in the calculations, during
the core He burning phase the luminosity is brighter, the lifetime is shorter
and the blue loop is less extended and rises less steeply than in absence
of overshooting  \citep[see e.g.][]{Matra82, ChinSto91}.

The possible
occurrence of convective shells and the consequent modification
of the internal chemical profile are strictly related to the adopted stability
criterion, either Schwarzschild or Ledoux \citep[see e.g.][]{SalBre99,
stot94, Canu00}. However the stability criterion has a marginal effect on
stellar models ranging from $\approx$3M$_{\odot}$ to $\approx$12 M$_{\odot}$,
as the ones taken into account in our work \citet{stot93}.

Moreover downward convecting overshooting from the outer convection zone
which modifies the envelope chemical profile 
influences the formation of blue loops.

For the same reason the occurrence of blue loops is thus
influenced by the mixing length value $\alpha$, which governs the convection
efficiency in the external convective envelope.
However when a blue loop develops (it is not true in some cases of small
and moderate $\alpha$ values) the maximum effective temperature during the
blue loop was found to be only a very weak function of $\alpha$ \citep[see
e.g.][]{ChinSto91}.
Moreover, since the atmospheric
model is fixed, $\alpha$ is generally calibrated through the
comparison between theory and observation for the stellar color. 
For this reason, in this paper we will not
include variations of this parameter, which is fixed to $\alpha$=1.9.

Another important phenomenon affecting the evolution of He burning
intermediate mass stars is mass loss \citep[see e.g. the review by][]{ChiMae86}
or the more recent work by \citet{SalBre99}.  As discussed by several authors
\citep[see e.g.][]{SalBre99} mass loss mainly influences the evolution of
masses higher than $\approx$15 M$_{\odot}$ but its influence on the loop
morphology, as discussed in Sec.~\ref{sec:massloss}, can be significant also for lower
masses.  However the physical mechanisms which affect mass loss and its
dependence on  metallicity are not still completely understood and thus the
range of possible efficiencies for the phenomenon is very wide \citep[see
e.g. the discussion in][]{MaeMey00, Vaz07}. In this paper we work in the canonical regime, also because we concentrate on intermediate mass stars but to take into account a possible mass loss contribution,
in Sec.~\ref{sec:massloss} we discuss some additional models computed with different mass loss efficiencies.

Due to the dependence of the loop extension on several physical
parameters, if one compares recent sets of intermediate mass stars models
available in the literature (which in general differ by the assumed chemical
composition, the adopted physical inputs, the external convection efficiency
etc..) one finds significant variations in the loop morphology.
In the following sections we will discuss the main parameters which have
already been demonstrated to potentially affect in a relevant way the blue
loop morphology: He burning cross sections, chemical composition, overshooting
and mass loss efficiency.

\section{Dependence of the loop extension on the uncertainties 
on the nuclear reaction cross sections}\label{sec:rates}

Several authors have investigated the dependence of the loop extension on
the H burning (through the CNO cycle) and the He burning reaction rates \citep[see
e.g.][]{bru90,renz92,sesti02,Xu2004a,weiss05}, and have shown that the
loop morphology is sensitive to variations of the adopted cross sections.

In this section we focus on the dependence of the loop formation and its
temperature extension on variations of the relevant nuclear fusion cross
sections within their estimated uncertainties. For stars in the central
He burning phase, the most relevant nuclear reactions are the 3$\alpha$, the
$^{12}$C($\alpha$,$\gamma$)$^{16}$O, that affects the He burning efficiency
and the $^{14}$N(p,$\gamma$)$^{15}$O that governs the shell
H burning through the CNO cycle.

\subsection{  $^{14}$N(p,$\gamma$)$^{15}$O cross section}

In stars with internal temperatures higher than $\approx 2\cdot 10^7$ K, the
hydrogen burning is dominated by the CNO cycle whose rate is
determined by the slowest process, the $^{14}$N(p,$\gamma$)$^{15}$O reaction.
For this reaction the Gamow peak is at $\approx$~30~keV;
thus the capture cross section must be known down to this energy
at which it has a very low value, preventing a direct laboratory measurement.
Instead, the cross sections are measured at higher energies and
then extrapolated, by means of the astrophysical S(E) factor, down
to the Gamow peak region.  The occurrence of resonance and resonance
tails complicates the extrapolation and it results in very large
uncertainties on the evaluated cross sections.   Recently, the
$^{14}$N(p,$\gamma$)$^{15}$O cross section has 
been measured by the LUNA collaboration  close to 70 keV \citep[see
  e.g.][]{14n,14n-bis,lem06}, leading to an improved estimate of the
S(E) factor at 
astrophysical energies ($S(0) = 1.61 \pm 0.08$ keV, about half of the value of the NACRE
compilation by \citet{nacre}). 
 
To quantify the effect of a variation of the $^{14}$N+p cross section, we
computed stellar models for several stellar masses, varying the $^{14}$N+p
reaction rate of the estimated uncertainty with respect to the LUNA value
(adopted in our standard models), models IX and X of
Table~\ref{table:casitrattati}. For all the investigated cases the effect on
the loop morphology of the present uncertainty on the
$^{14}$N(p,$\gamma$)$^{15}$O rate is negligible, suggesting that this
uncertainty is not expected to contribute significantly to the final error
budget for the predicted Cepheid pulsation properties.

\subsection{ 3$\alpha$ cross section}

Recently \citet{3alpha-bis} reported new measurements for the 3$\alpha$ rate,
finding, in particular, a resonance at energies of $\approx$11 MeV;
however, in the temperature range of interest, the differences with
respect to the most widely adopted 3$\alpha$ rate are within the
uncertainty evaluated by the NACRE \citep{nacre} compilation, reaching a
maximum of $\approx$20\% at temperatures of about $10^8$ K \citep[see e.g.][]{weiss05}.  

For this reason we decided to take the error quoted by NACRE (20\%), in the
temperature range of interest, as a conservative estimate of the uncertainty on
the 3$\alpha$ reaction rate. We computed specific models by varying the
3$\alpha$ rate within the estimated uncertainty for different stellar masses
(in our ``standard'' case) finding no relevant variations in the loop
extension. These results suggest  that the present uncertainties on the 3$\alpha$
rate do not substantially influence the evolutionary and pulsational
quantities for Cepheids stars.

\subsection{$^{12}$C($\alpha$,$\gamma$)$^{16}$O cross section}

The measurement of the $^{12}\mathrm{C}+\alpha$ cross section at the energies
of interest in stars ($\approx$ 300 keV) is presently impossible due to the
smallness of the predicted value ($\simeq 10^{-17}$ b).  The extrapolation
from higher energies is quite difficult, because the cross section in this
energy region is a mixture of groundstate and cascade transitions, and the
cross section for the ground state transitions is dominated by the tails of
subthreshold resonances with the interference of other processes \citep[see
e.g.][]{3alpha}.  For these reasons the uncertainty quoted in the past for this rate 
was very large, up to a factor of two, and the effect of this error on the loop
extension was very strong \citep[see e.g.][]{bru90}.  Given the astrophysical
importance of this reaction, in the past decade a significant experimental
effort has been devoted to the improvement of the determination of its cross
section.  Recent results by \citet{12c} of direct
$^{12}$C($\alpha$,$\gamma$)$^{16}$O measurements in the temperature range 0.001$\le T_9 \le$ 10, $T_9$ being the temperature in $10^9$ K, lead to a new determination of the extrapolated astrophysical
reaction rate with a maximum total uncertainty of $\pm$ 25\%
\citep[see][]{12c}.

In the present work we adopt for the $^{12}$C($\alpha$,$\gamma$)$^{16}$O reaction rate
the parametrization and the parameter values given by
\citet{12c}; at the typical He burning temperature ($10^8 \div 2 \cdot 
10^8$ K) this new rate is about 20\% lower than the NACRE rate.

As discussed in \citet{bru90}, the blue loop extension
depends on the $^{12}\mathrm{C}+\alpha$ cross section: the higher the
rate of this reaction, the higher the effective temperature excursion of
the loop.  Several authors \citep[see e.g.][]{sesti02,Xu2004a}  showed that the
$^{12}$C($\alpha$,$\gamma$)$^{16}$O cross section is related to the efficiency
of the envelope contraction and thus to the loop extension.\\
We evaluated the influence on the loop extension of the present uncertainty on the
$^{12}$C($\alpha$,$\gamma$)$^{16}$O reaction rate for
models of different masses. To this aim we calculated two additional sets 
of tracks with ``standard'' chemical composition (Z=0.005 Y=0.26) 
and the $^{12}$C+$\alpha$ rate by  \citet{12c} increased or decreased 
by the estimated uncertainty of 25\% (models VII and VIII of Table~\ref{table:casitrattati}).
As shown in Fig.~\ref{fig:12C}, available online, the effect
is negligible for all the masses.

\onlfig{1}{
\begin{figure}
\bigskip
\centering
\includegraphics[width=6.5cm]{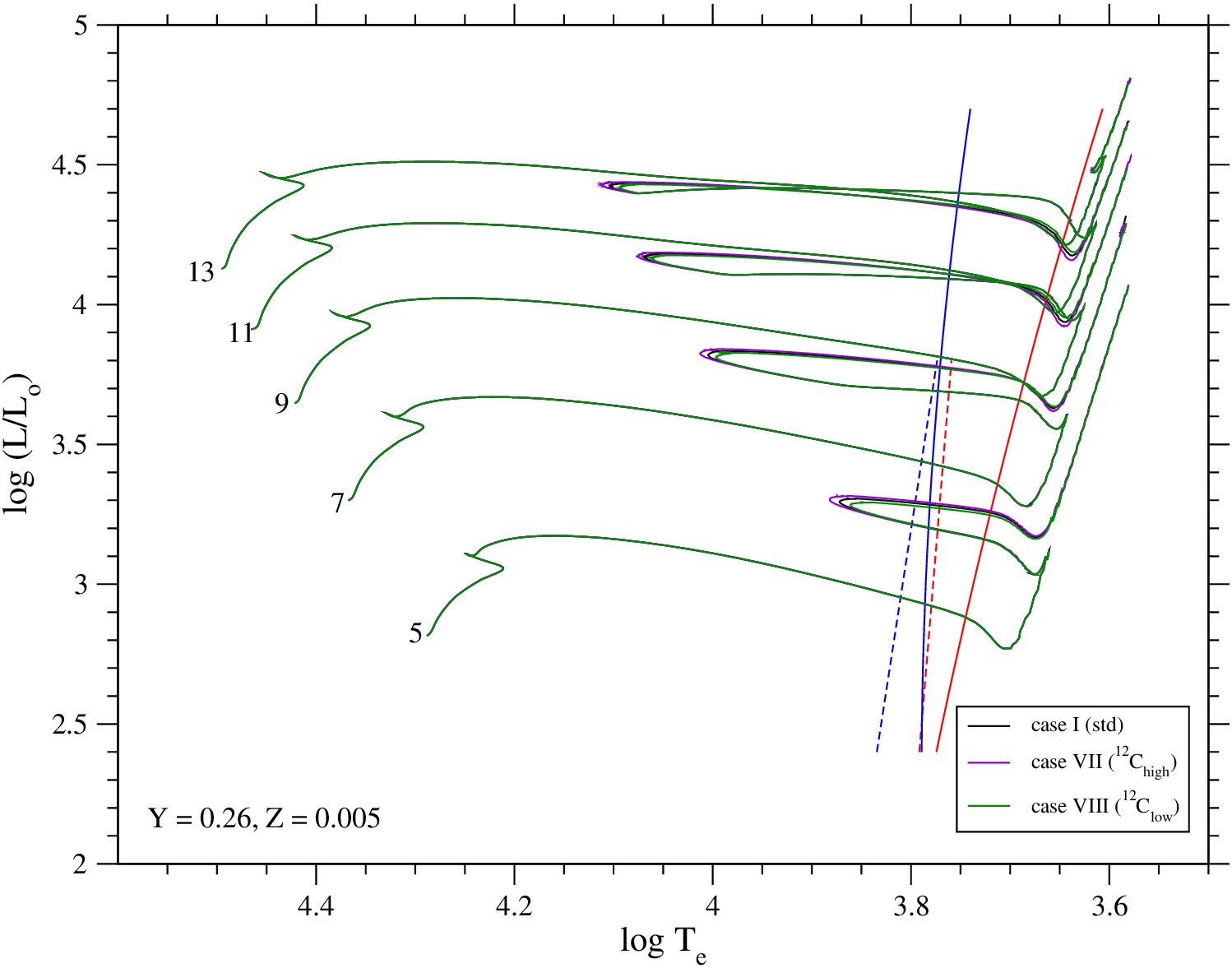}\hspace{2mm}
\caption{Evolutionary tracks with standard chemical composition
with three different values of the $^{12}$C($\alpha$,$\gamma$)$^{16}$O
rate (see text).} 
\label{fig:12C} 
\end{figure}
}

Thus, at variance with results published in the past, now
the uncertainty in the $^{12}$C+ $\alpha$ nuclear rate is noticeably reduced
so that its effect on the loop morphology appears negligible.  Consequently, in this paper, we neglect the uncertainty associated with this cross
section in the evaluation of the error budget.

In conclusion, if one takes at their face value the present estimated
uncertainties of the cross sections relevant for Cepheids star evolution, one
finds that now the precision of the experimental measurements, at the energies
of astrophysical interest, is high enough to rule out significant effects of
these uncertainties on the loop extension and thus on the pulsational
quantities.

\section{Effects of the chemical composition uncertainties}\label{sec:deltaz}

In this section we present the effects of the LMC Cepheid chemical
composition uncertainties on evolutionary properties such as the He burning
loop extension and the $ML$ relation, and on pulsation properties such as the
instability strip morphology and the $PL$ relations.

For the LMC iron content we adopt [Fe/H]=-0.40 that, as quoted above, is
consistent with the estimates by \citet{luck98} and \citet{r08}. In particular
\citet{luck98} suggest a range of -0.55$\leq$[Fe/H]$\leq$-0.19. By adopting
the \citet{aspl05} solar mixture, that is a $(Z/X)_{\odot}$ value of 0.0165,
from the assumed [Fe/H] one obtains (see Eq.~\ref{eq:zeta}) $Z=0.005$ (our
``standard'' value). On the basis of the observed range suggested by
\citet{luck98} we allowed $Z$ to vary in the range 0.0035$\leq Z \leq$0.008.
Taking into account the current uncertainties on $\Delta{Y}/\Delta{Z}$
\citep[see e.g.][]{papo98,gen08}, we allowed a variation of the original
helium content from $Y$=0.26 to $Y$=0.28. With $Z=0.005$, $Y=0.26$ corresponds to
$\Delta{Y}/\Delta{Z}$ = 2, while $Y=0.28$ implies $\Delta{Y}/\Delta{Z}$ = 6.

One should also take into account the uncertainty on the solar mixture.  As
already discussed, tridimensional hydrodynamical atmospheric models by
\citet{aspl05} reduced the derived abundances of CNO and other heavy 
elements with respect to previous estimates \citep[][hereafter
  GS98]{gre98}. GS98 already improved the mixture by \citet{gre93},
widely adopted in the literature, mainly revising the 
CNO and Ne abundance and confirming the very good agreement between the new
photospheric and meteoric results for the solar iron abundance.

The change of the adopted heavy element ratios might affect the evolutionary calculations
of LMC Cepheids in two different ways: 1) changing the evolutionary
characteristics at fixed metallicity; 2) changing the LMC inferred
metallicity from the observed [Fe/H]. Regarding the second point, the adoption
of the GS98 solar mixture leads to a LMC metallicity of $Z \approx$0.008, which
is within the previous quoted metallicity uncertainty. The first point
is analysed by taking into account the effects on the evolutionary features of a
solar mixture variation at fixed metallicity, as due
to its effect on opacity, nuclear burning and equation of state
calculations. 
 
In the following we present different sets of models varying the  chemical composition within the quoted ranges of uncertainty. The combinations of $Y$ and $Z$ values
adopted for the different sets of models are summarized in
Table~\ref{table:casitrattati}.  The first column of
Table~\ref{table:casitrattati} gives the reference name of the treated
cases ($std$, $Z_{low}$, $Z_{high}$ etc..), the second and the third columns
give the adopted $Y$ and $Z$ values, whereas the fourth column reports the
adopted chemical mixture at fixed $Z$. The cases XI and XII will be presented
in detail in Subsec.\ref{oversh} in which we will discuss the effect of a mild
core overshooting in the H-burning phase.

For each case of Table~\ref{table:casitrattati} we computed evolutionary and
nonlinear convective pulsation models spanning a stellar mass range from
3-4M$_{\odot}$ to 13-14M$_{\odot}$, which covers the range of masses
crossing the instability strip (see for details Sec.\ref{sec:ML}).

\begin{table}
\caption{Computed sets of evolutionary tracks and pulsation
  models. GS98 = \citet{gre98}, AGS05 = \citet{aspl05}.}      
\label{table:casitrattati}   
\centering                     
\begin{tabular}{l l l l}       
\hline\hline
Case & $Y$ & $Z$ & Mixture \\  
\hline  
I  ($std$)   & 0.26  & 0.005  & AGS05 \\ 
II ($Z_{low}$)  & 0.26  & 0.0035 & AGS05 \\
III ($Z_{high}$)  & 0.26  & 0.008  & AGS05 \\
IV ($Y_{high}$)   & 0.28  & 0.005  & AGS05 \\
V ($Y_{high}$, $Z_{high}$)&0.28  & 0.008  & AGS05 \\
\hline
VI ($Z_{high}$, GS98)&0.26& 0.008  &GS98 \\
\hline
VII  ($^{12}$C($\alpha$,$\gamma$)$^{16}$O)$_{high}$   &0.26 &0.005&AGS05\\ 
VIII ($^{12}$C($\alpha$,$\gamma$)$^{16}$O)$_{low}$  &0.26 &0.005&AGS05\\ 
IX ($^{14}$N(p,$\gamma$)$^{15}$O)$_{high}$  &0.26 &0.005&AGS05\\ 
X ($^{14}$N(p,$\gamma$)$^{15}$O)$_{low}$  &0.26 &0.005&AGS05\\ 
\hline
XI ($l_{\rm ov}=0.1 H_{\rm p}$) &0.26 &0.005&AGS05 \\ 
XII ($l_{\rm ov}=0.25 H_{\rm p}$) &0.28 &0.005&AGS05\\
XIII ($l_{\rm ov}=0.25 H_{\rm p}$) &0.26 &0.005&AGS05\\
\hline                                
\end{tabular}
\end{table}
\noindent
All the computed sets for the cases I to VI of
Table~\ref{table:casitrattati} 
are plotted in Fig.~\ref{fig:set-tracce}; the fits of the corresponding
instability strips calculated as explained in Subsec.\ref{sec:strip}
are superimposed on the evolutionary models.\\
Some models show a smaller secondary loop
during the redward excursion in the central He burning phase \citep[see
e.g.][too]{B00,SalBre99}, which is not a numerical artifact. The physical
reason for this secondary loop is a slight increase in the size of the 
convective
core, due to the opacity increase in the C and O rich nucleus, leading to
small variations of the H and He burning efficiency. The mechanism is the
analog of the breathing pulses occurring in low mass stars \citep[see
e.g.][]{Swei72, Swei73,Cast85}.

\subsection{Loop extension}\label{sec:loop}

\begin{figure}
\bigskip
\centering
\includegraphics[width=6.5cm]{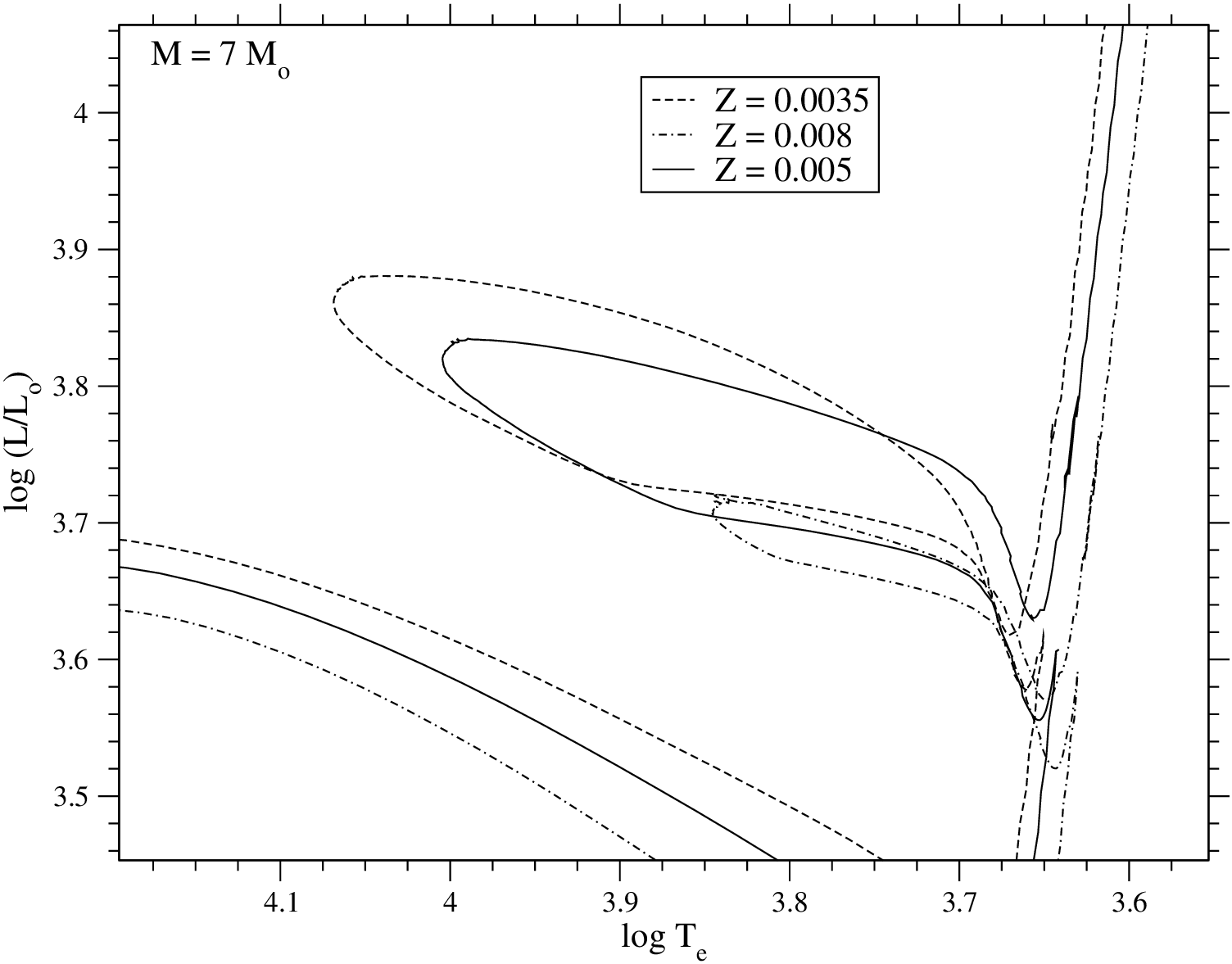}\vspace{4mm}
\includegraphics[width=6.5cm]{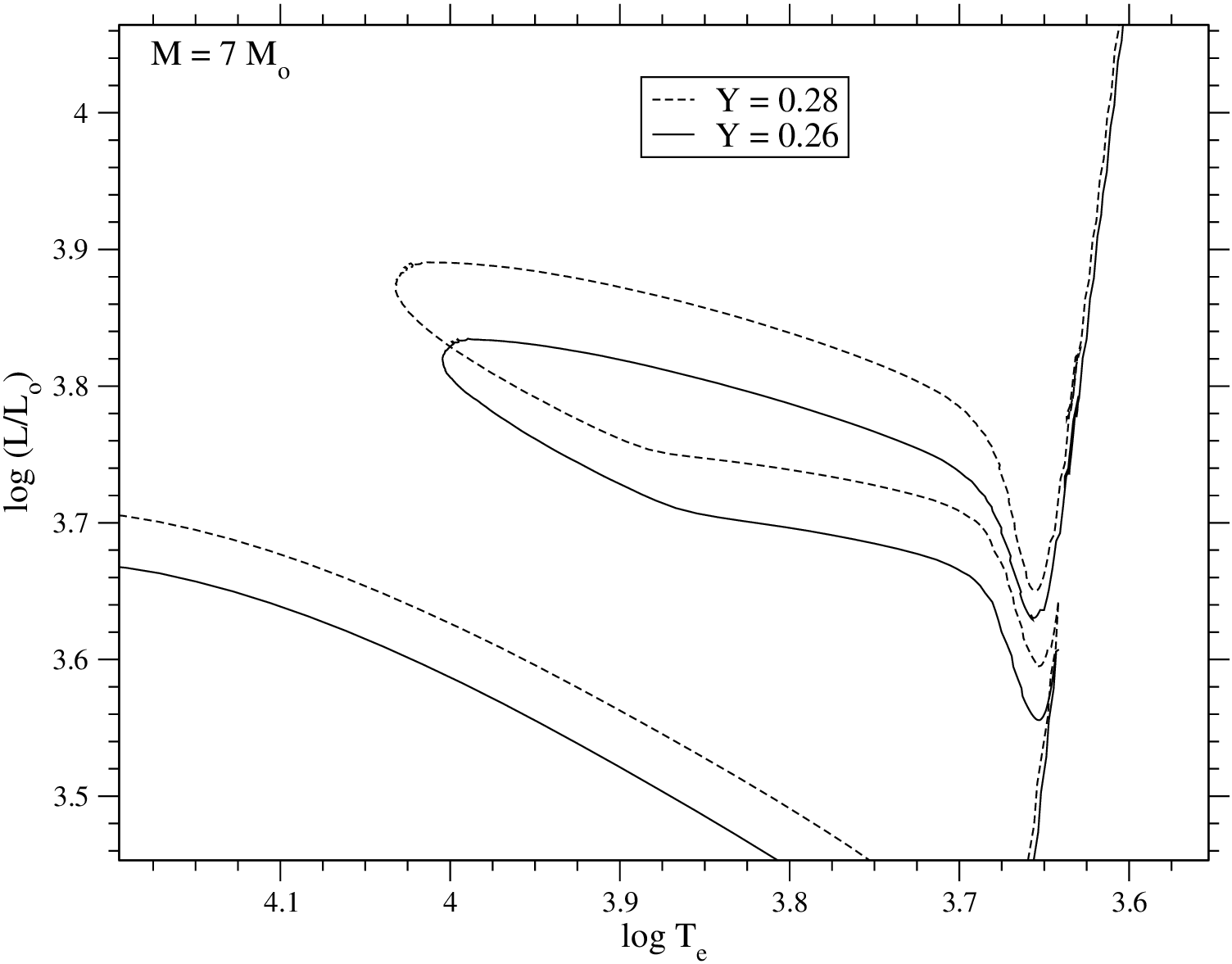}
\caption{Upper panel: effect of a metallicity variation on the loop extension for a 7 M$_{\odot}$; the three different lines correspond to cases I,II,III of Table~\ref{table:casitrattati}: $std$ case ($Z$=0.005, solid line)
$Z_{low}$=0.0035 (dashed line) and $Z_{high}$=0.008 (dot-dashed line). Lower panel: as in
the upper panel but for a helium abundance variation and for cases I ($std$, $Y$=0.26, solid line) and IV ($Y_{high}$=0.28, dashed line) of Table 1.}\label{fig:loop-DeltaZ} 
\end{figure}

Figure~\ref{fig:loop-DeltaZ} shows, for a 7 M$_{\odot}$ model, the
effects on the blue loop extension of variations in the helium and metal contents within the uncertainties of these quantities (cases from I to IV of
Table~\ref{table:casitrattati}). The upper panel shows that increasing the metal content, at fixed helium abundance, the extension of the blue loop decreases, whereas the lower panel shows that an increase of the helium content within its
uncertainty range, at fixed metallicity, leads to more extended loops, even if the effect is reduced compared to that of lowering $Z$.
The behaviour is similar for the other analysed model masses.

A simultaneous variation of $Z$ and $Y$ within their uncertainty
ranges does not result in an additive effect on the temperature extension of
the loop. An example of this behaviour is shown in Fig.~\ref{fig:set-tracce}
(f), for which the set of evolutionary tracks was computed with the upper values
of $Y$ and $Z$, namely $Y$=0.28 and $Z$=0.008 (case V of Table~\ref{table:casitrattati}). Beside the apparent non additivity
 in
$\log T_{e}$, there is rather complicated mass dependence with a more regular trend
for the intermediate mass values ($\simeq 5-8 M_\odot$) and an inverted trend
for higher masses. 
Thus one should be very careful to predict the extension of the loop when more
than one parameter is varied, as it cannot be simply evaluated 
by combining the results obtained for the variation of each single parameter separately.

The effect of changing the mixture at fixed $Z$ is shown by comparing
Fig.~\ref{fig:set-tracce} (c) (case III of Table~\ref{table:casitrattati}), in
which the set of tracks is computed with $Z$ = 0.008 and the mixture of AGS05,
with Fig.~\ref{fig:set-tracce} (e) (case VI of Table~\ref{table:casitrattati}),
computed with the same total metallicity ($Z$ = 0.008) 
but with  the GS98 mixture. To analyse this
case we changed the original element abundances of our models and we
calculated suitable opacity tables making advantage of the codes available at
the OPAL web site\footnote{http://www-phys.llnl.gov/Research/OPAL/opal.html};
the EOS is also varied following the mixture change \citep[see][for a more detailed analysis of the effects of a mixture
update on the stellar models]{deg06}.\\
The mixture change at fixed $Z$ has a negligible effect on the loop extension
for $M \le 8 M_{\odot}$.  For higher masses we find that models with \citet{aspl05}
composition do not show blue loops at all, whereas the ones with the 
\citet{gre98} mixture display extended blue loops. As this effect is evident 
only for about half of the analyzed mass range we decided to neglect it in 
the following.

\subsection{Instability strip}\label{sec:strip}

For each chemical composition, mass and luminosity, the modal stability of pulsation models has been investigated for both the fundamental and the first
overtone mode. The effective temperatures (hereafter in K) of the hottest and coolest pulsating
fundamental models, increased and decreased, respectively, by the extension of
the pass in temperature adopted in our analysis, 50 K, correspond to the
fundamental blue and red edges (FBE and FRE). Similarly the first overtone
blue (FOBE) and red (FORE) edges are evaluated. Table~\ref{table:fitparabolici}
reports the quadratic fit coefficients of the FBE and FRE, plotted in
Fig.~\ref{fig:set-tracce} as solid lines.  The linear fits of the FBE and FRE
are available on line in Table~\ref{table:fitlineari}, whereas the linear fits
for the FOBE and FORE, plotted in Fig.~\ref{fig:set-tracce} as dashed lines,
are reported in Table~\ref{table:fitlineariFO} in the form $\log T_e = a + b\; \log (L/L_{\odot})$, where $\sigma$ is the least square estimate of the intrinsic dispersion. Figures~\ref{fig:set-tracce}
(a)-(f) show the evolutionary tracks superimposed with the quadratic fits of
the edges of the fundamental mode instability strips and the linear fits of
the first overtone ones for the cases I,II,III,IV,V,VI of
Table~\ref{table:casitrattati}.

\onltab{2}{
\begin{table}
\caption{Coefficients of the linear analytical relations for the fundamental red and blue edges.}  
\label{table:fitlineari}  
\centering 
\begin{tabular}{l l l l }       
\hline\hline
Case & $a$ & $b$ & $\sigma$\\  
\hline
Blue Edge\\  
 \hline
I  ($std$)      & 3.848 $\; \pm \;$ 0.010 & -0.022  $\; \pm \;$ 0.002 & 0.005 \\
II ($Z_{low}$)     & 3.828 $\; \pm \;$ 0.013 & -0.016  $\; \pm \;$ 0.003 & 0.006 \\
III ($Z_{high}$)     & 3.853 $\; \pm \;$ 0.029 & -0.025  $\; \pm \;$ 0.008 & 0.012 \\
IV ($Y_{high}$)      & 3.866 $\; \pm \;$ 0.020 & -0.026  $\; \pm \;$ 0.005 & 0.009 \\
V ($Y_{high}$, $Z_{high}$)  & 3.870 $\; \pm \;$ 0.020 & -0.029  $\; \pm \;$ 0.006 & 0.008 \\
VI ($Z_{high}$, GS98)& 3.858 $\; \pm \;$ 0.022 & -0.027  $\; \pm \;$ 0.007 & 0.011 \\
\hline    
Red Edge\\ 
 \hline 
I  ($std$)      & 3.955  $\; \pm \;$ 0.020 & -0.073  $\; \pm \;$ 0.005 & 0.010 \\
II ($Z_{low}$)     & 3.922  $\; \pm \;$ 0.048 & -0.061  $\; \pm \;$ 0.012 & 0.023 \\
III ($Z_{high}$)     & 3.974  $\; \pm \;$ 0.028 & -0.082  $\; \pm \;$ 0.008 & 0.012 \\
IV ($Y_{high}$)      & 3.934  $\; \pm \;$ 0.021 & -0.066  $\; \pm \;$ 0.005 & 0.009 \\ 
V ($Y_{high}$, $Z_{high}$)  & 3.965  $\; \pm \;$ 0.019 & -0.078  $\; \pm \;$ 0.005 & 0.008 \\
VI ($Z_{high}$, GS98)& 3.958  $\; \pm \;$ 0.018 & -0.074  $\; \pm \;$ 0.005 & 0.009 \\
\hline                           
\end{tabular}
\end{table}
}

\begin{table*}
\bigskip
\caption{The same as Table 2 but for quadratic fits.}      
\label{table:fitparabolici}  
\centering 
\begin{tabular}{l l l l l}       
\hline\hline
Case & $a$ & $b$ & $c$ & $\sigma$\\  
\hline
Fundamental blue edge\\ 
 \hline 
I  ($std$)      & 3.744 $\; \pm \;$ 0.020  & 0.039 $\; \pm \;$ 0.012 & -0.009 $\; \pm \;$ 0.002 & 0.002 \\
II ($Z_{low}$)     & 3.751 $\; \pm \;$ 0.058  & 0.028 $\; \pm \;$ 0.032 & -0.006 $\; \pm \;$ 0.004 & 0.005 \\
III ($Z_{high}$)     & 3.594 $\; \pm \;$ 0.060  & 0.137 $\; \pm \;$ 0.037 & -0.024 $\; \pm \;$ 0.005 & 0.005 \\
IV ($Y_{high}$)      & 3.672 $\; \pm \;$ 0.032  & 0.086 $\; \pm \;$ 0.018 & -0.016 $\; \pm \;$ 0.003 & 0.003 \\
V ($Y_{high}$, $Z_{high}$)  & 3.748 $\; \pm \;$ 0.104  & 0.046 $\; \pm \;$ 0.063 & -0.011 $\; \pm \;$ 0.009 & 0.008 \\
VI ($Z_{high}$, GS98)& 3.599 $\; \pm \;$ 0.076  & 0.131 $\; \pm \;$ 0.046 & -0.023 $\; \pm \;$ 0.007 & 0.007 \\
\hline    
Fundamental red edge\\  
 \hline 
I  ($std$)      & 3.878  $\; \pm \;$ 0.112  &-0.028 $\; \pm \;$ 0.064 & -0.006 $\; \pm \;$ 0.009 & 0.010  \\
II ($Z_{low}$)     & 4.071  $\; \pm \;$ 0.261  &-0.146 $\; \pm \;$ 0.147 &  0.012 $\; \pm \;$ 0.020 & 0.025  \\
III ($Z_{high}$)     & 3.728  $\; \pm \;$ 0.059  & 0.071 $\; \pm \;$ 0.036 & -0.023 $\; \pm \;$ 0.005 & 0.004  \\
IV ($Y_{high}$)      & 3.988  $\; \pm \;$ 0.118  &-0.098 $\; \pm \;$ 0.068 &  0.004 $\; \pm \;$ 0.009 & 0.010  \\
V ($Y_{high}$, $Z_{high}$)  & 3.788  $\; \pm \;$ 0.032  & 0.031 $\; \pm \;$ 0.019 & -0.016 $\; \pm \;$ 0.003 & 0.002  \\
VI ($Z_{high}$, GS98)& 3.731  $\; \pm \;$ 0.031  & 0.064 $\; \pm \;$ 0.019 & -0.020 $\; \pm \;$ 0.003 & 0.003  \\
\hline                           
\end{tabular}
\end{table*}

\begin{table}
\caption{The same as Table 2 but for the first overtone boundaries.}      
\label{table:fitlineariFO}  
\centering 
\begin{tabular}{l l l l }       
\hline\hline
Case & $a$ & $b$ & $\sigma$\\  
\hline
First overtone  blue edge\\  
 \hline
I  ($std$)      & 3.939 $\; \pm \;$ 0.019 & -0.043 $\; \pm \;$ 0.006 & 0.005 \\
II ($Z_{low}$)     & 3.942 $\; \pm \;$ 0.019 & -0.044 $\; \pm \;$ 0.006 & 0.005 \\
III ($Z_{high}$)     & 3.913 $\; \pm \;$ 0.046 & -0.038 $\; \pm \;$ 0.015 & 0.013 \\
IV ($Y_{high}$)      & 3.922 $\; \pm \;$ 0.027 & -0.038 $\; \pm \;$ 0.008 & 0.007 \\
V ($Y_{high}$, $Z_{high}$)  & 3.954 $\; \pm \;$ 0.012 & -0.050 $\; \pm \;$ 0.004 & 0.003 \\
VI ($Z_{high}$, GS98)& 3.962 $\; \pm \;$ 0.012 & -0.052 $\; \pm \;$ 0.004 & 0.005 \\
\hline    
First overtone red edge\\ 
 \hline 
I  ($std$)      & 3.847 $\; \pm \;$ 0.010 & -0.023 $\; \pm \;$ 0.003 & 0.003 \\
II ($Z_{low}$)     & 3.834 $\; \pm \;$ 0.004 & -0.018 $\; \pm \;$ 0.001 & 0.001 \\
III ($Z_{high}$)     & 3.840 $\; \pm \;$ 0.009 & -0.024 $\; \pm \;$ 0.003 & 0.003 \\
IV ($Y_{high}$)      & 3.834 $\; \pm \;$ 0.004 & -0.018 $\; \pm \;$ 0.001 & 0.001 \\
V ($Y_{high}$, $Z_{high}$)  & 3.826 $\; \pm \;$ 0.004 & -0.018 $\; \pm \;$ 0.001 & 0.001 \\
VI ($Z_{high}$, GS98)& 3.831 $\; \pm \;$ 0.014 & -0.021 $\; \pm \;$ 0.005 & 0.006 \\
\hline                     
\end{tabular}
\end{table}

\begin{figure}
\bigskip
\centering
\includegraphics[width=6.5cm]{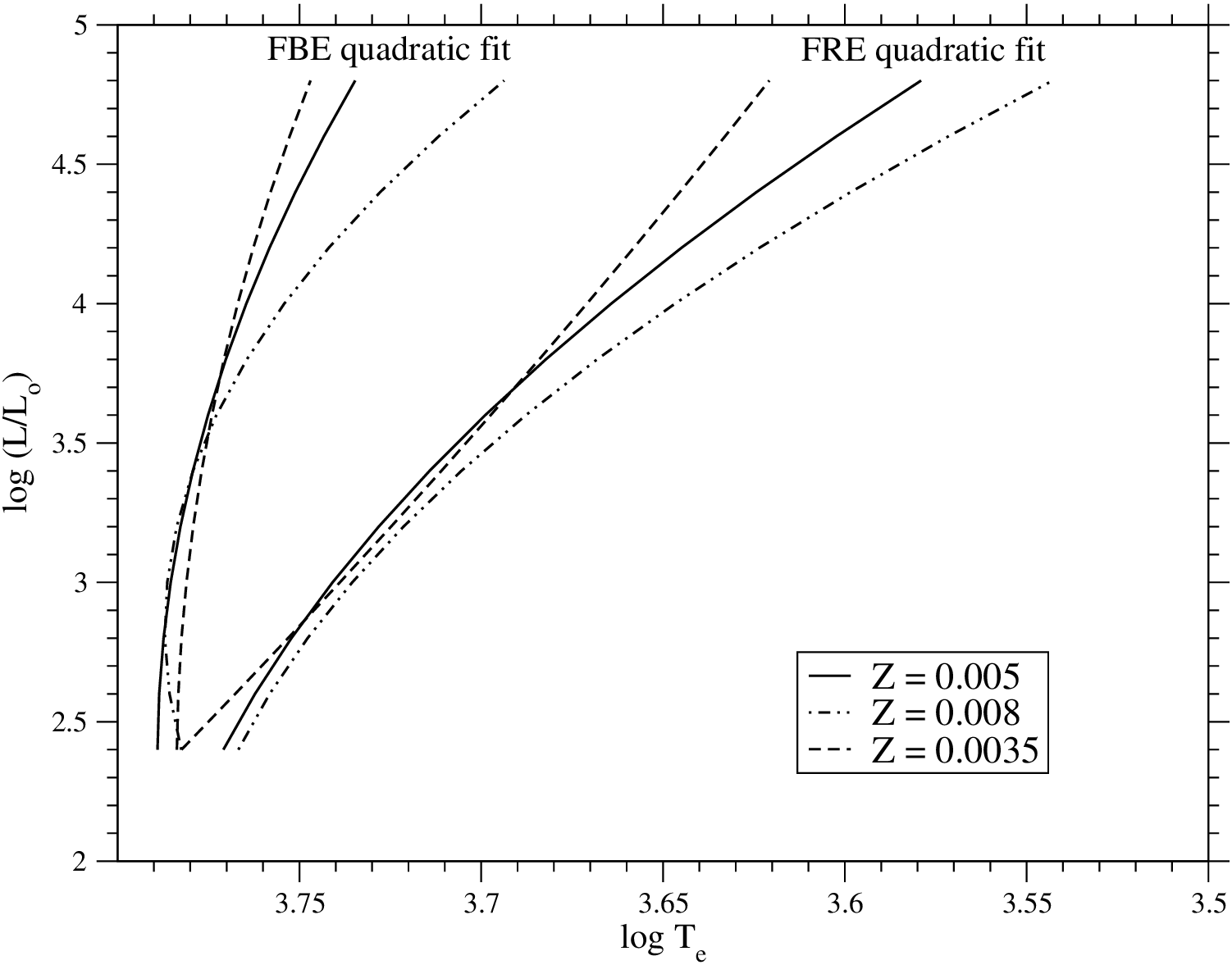}\vspace{4mm}
\includegraphics[width=6.5cm]{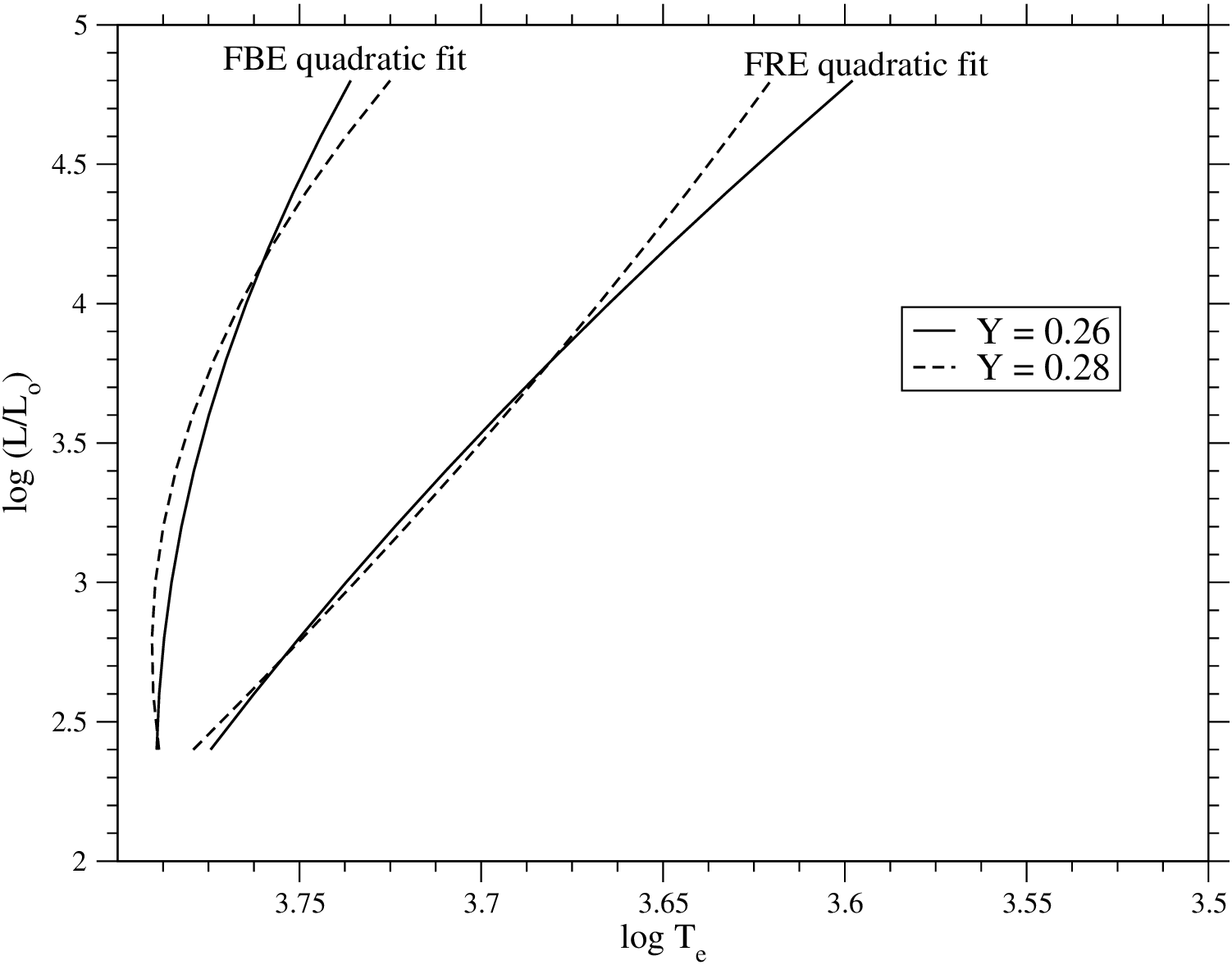}
\caption{Upper panel: effect on the fundamental instability strip edges of a metallicity
  variation within its uncertainty (cases I,II,III of
  Table~\ref{table:casitrattati}). Lower panel: as in the upper panel but for a variation
  of the original helium abundance (cases I, IV of Table~\ref{table:casitrattati}). In
  both panels the quadratic fits corresponding to Table~\ref{table:fitparabolici} are shown.}
\label{fig:strip-deltaZ} 
\end{figure}

To analyse the effect of the chemical composition uncertainty, the upper panel of
Fig.~\ref{fig:strip-deltaZ} shows the instability strips computed for the
different metallicities corresponding to the cases I, II, III of
Table~\ref{table:casitrattati}; whereas in the lower panel of the same figure
the effect on the instability strip of the $Y$ variation within its uncertainty is shown (cases I and
IV of Table~\ref{table:casitrattati}). 
As already found in our previous papers \citep{B99,cmm00} the
predicted instability strip becomes redder as the metallicity increases,
while increasing the helium content from 0.26 to 0.28 at $Z=0.005$ only
produces a slight narrowing of the instability strip at the highest
luminosity levels, likely due to the slightly reduced efficiency of
the H ionization region in driving pulsation. 

\subsection{The mass-luminosity relation}\label{sec:ML}

\begin{figure*}
\centering
\includegraphics[width=6.5cm]{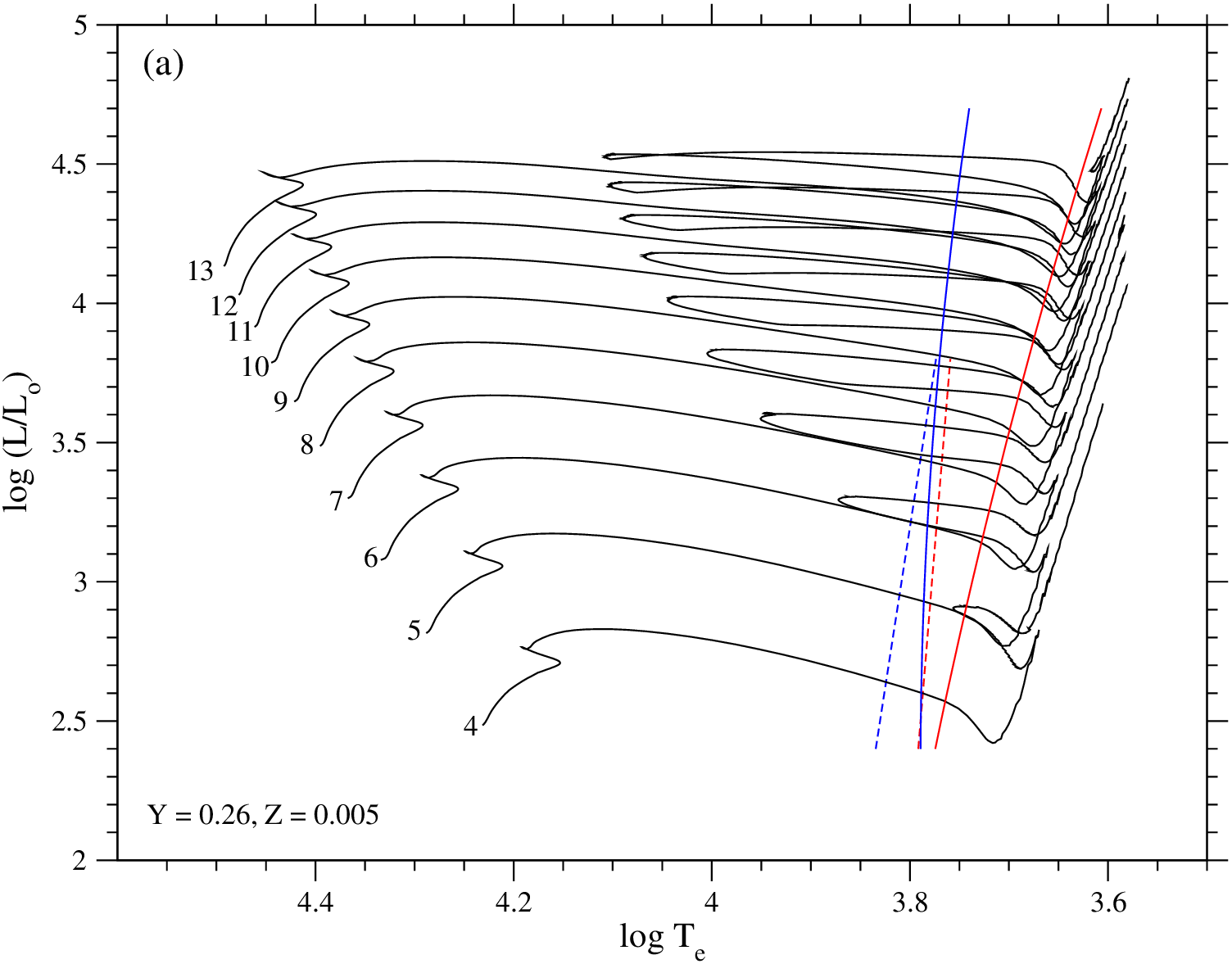}\hspace{2mm} 
\includegraphics[width=6.5cm]{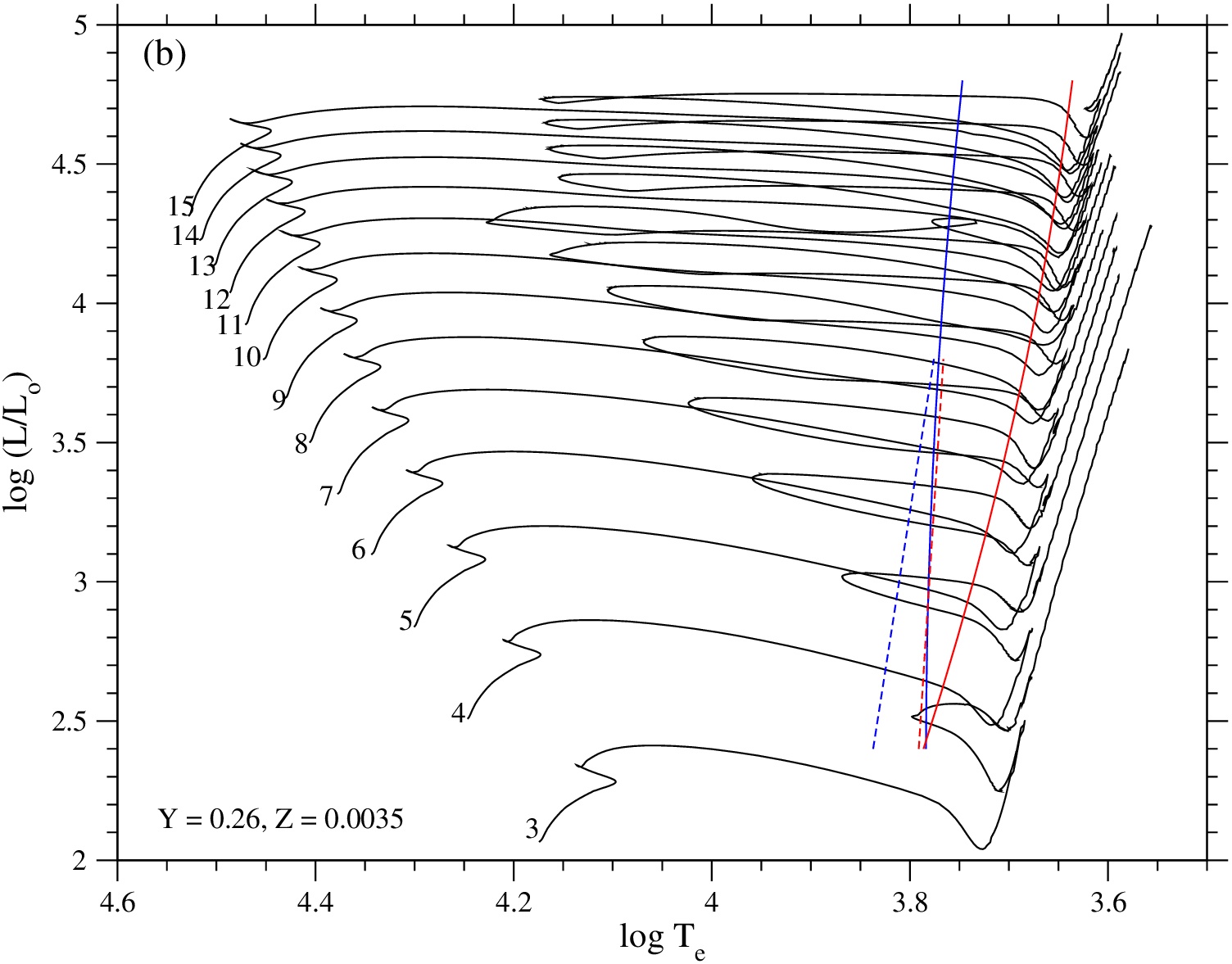}\vspace{2mm}\\
\includegraphics[width=6.5cm]{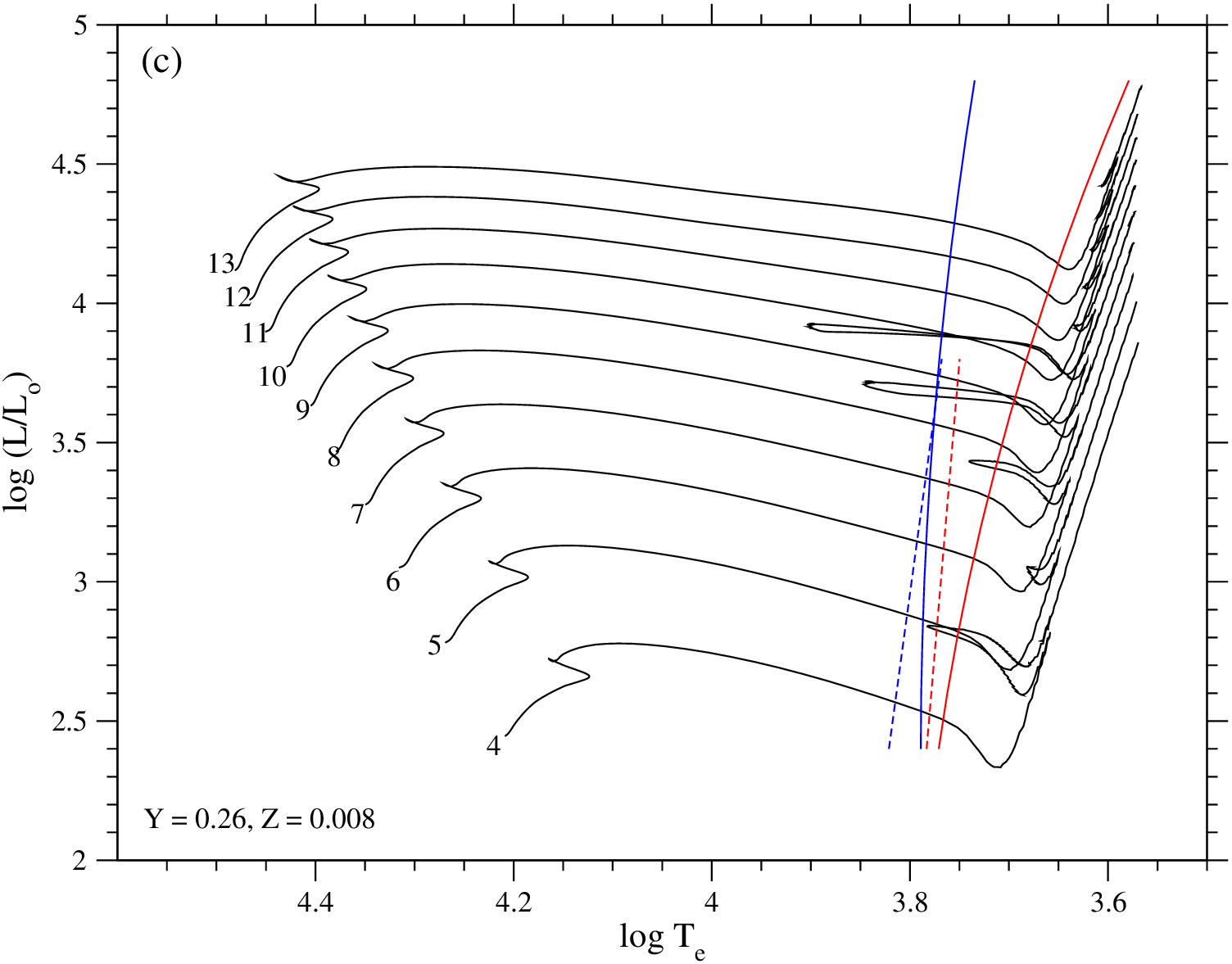}\hspace{2mm}
\includegraphics[width=6.5cm]{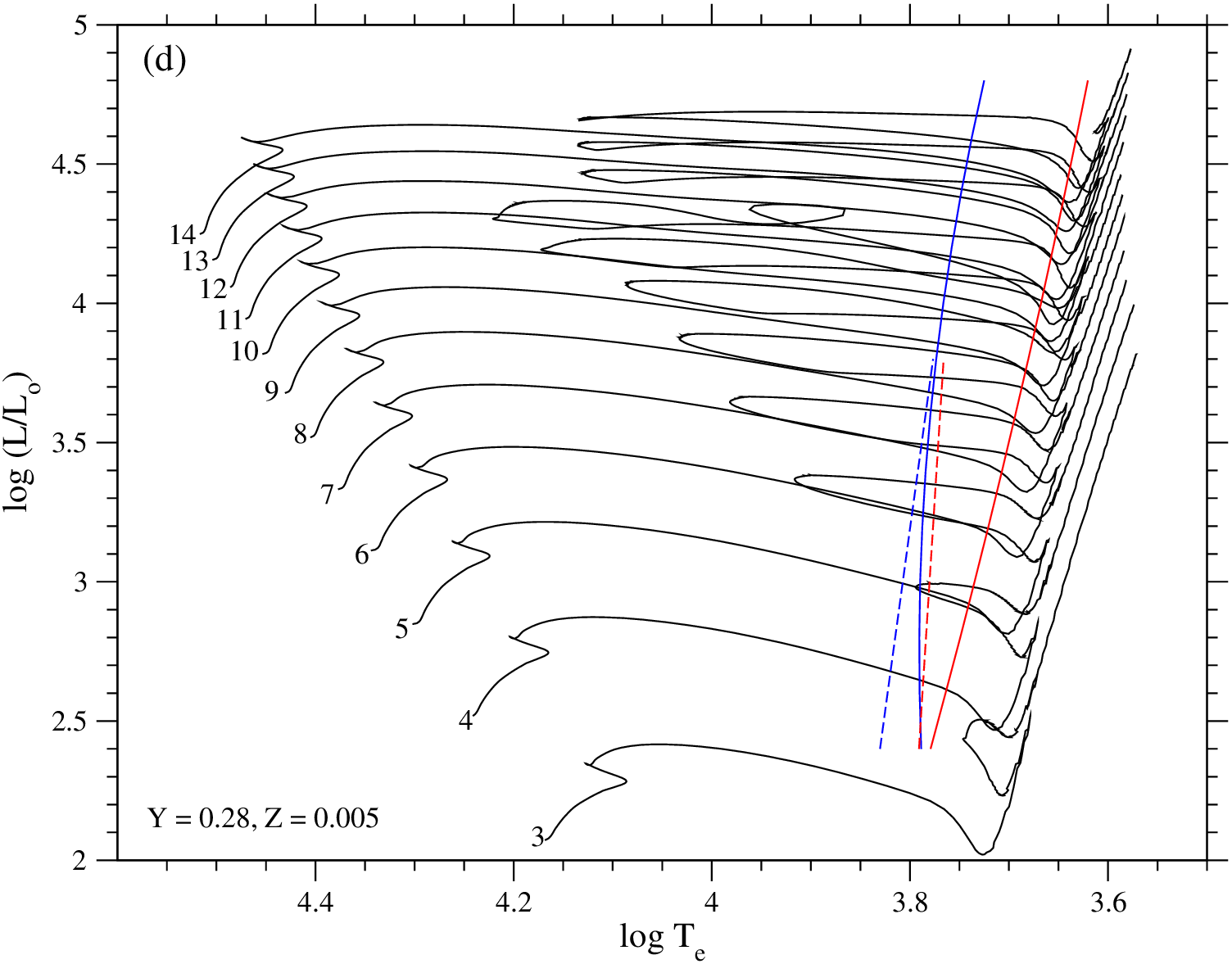}\vspace{2mm}\\
\includegraphics[width=6.5cm]{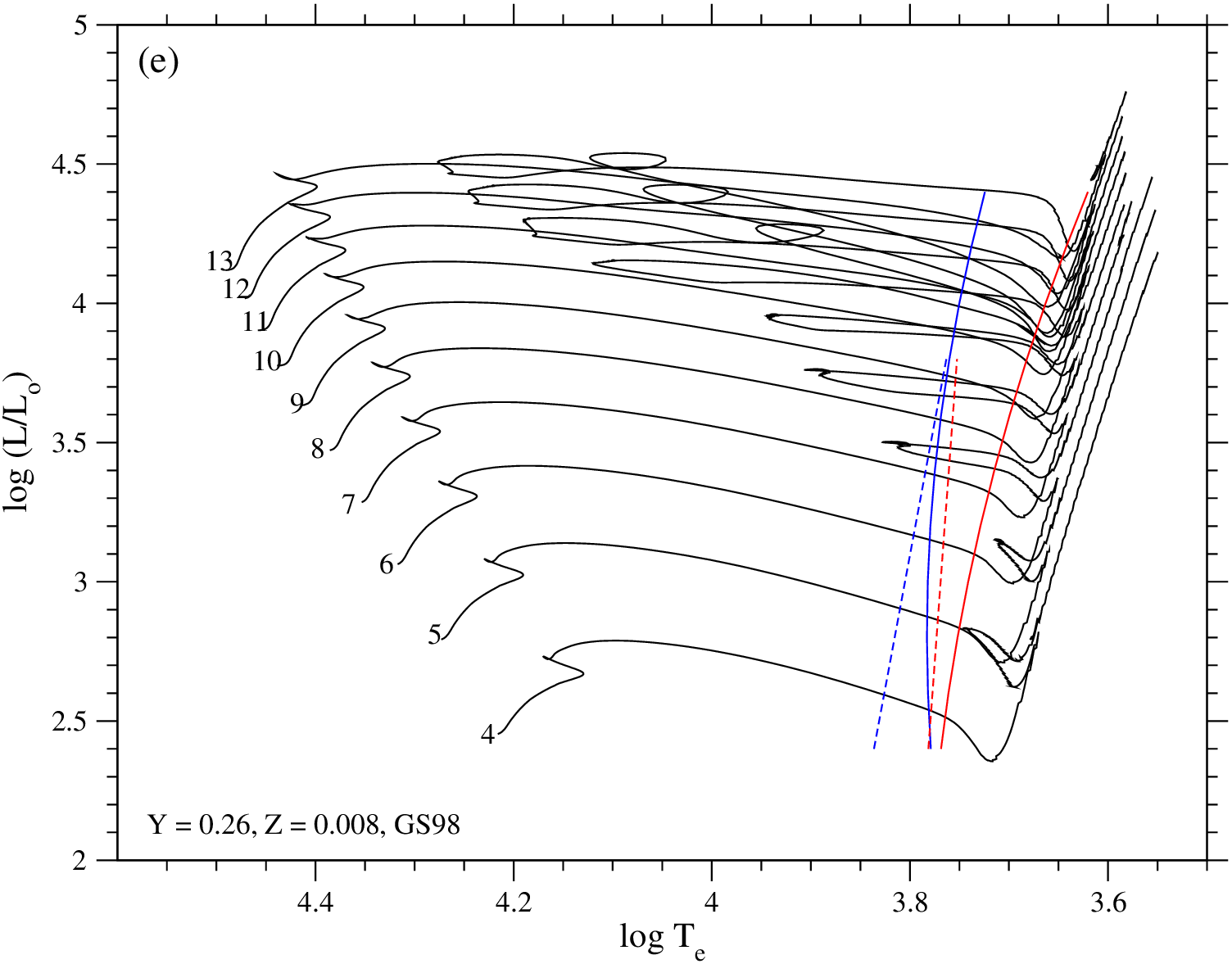}\hspace{2mm}
\includegraphics[width=6.5cm]{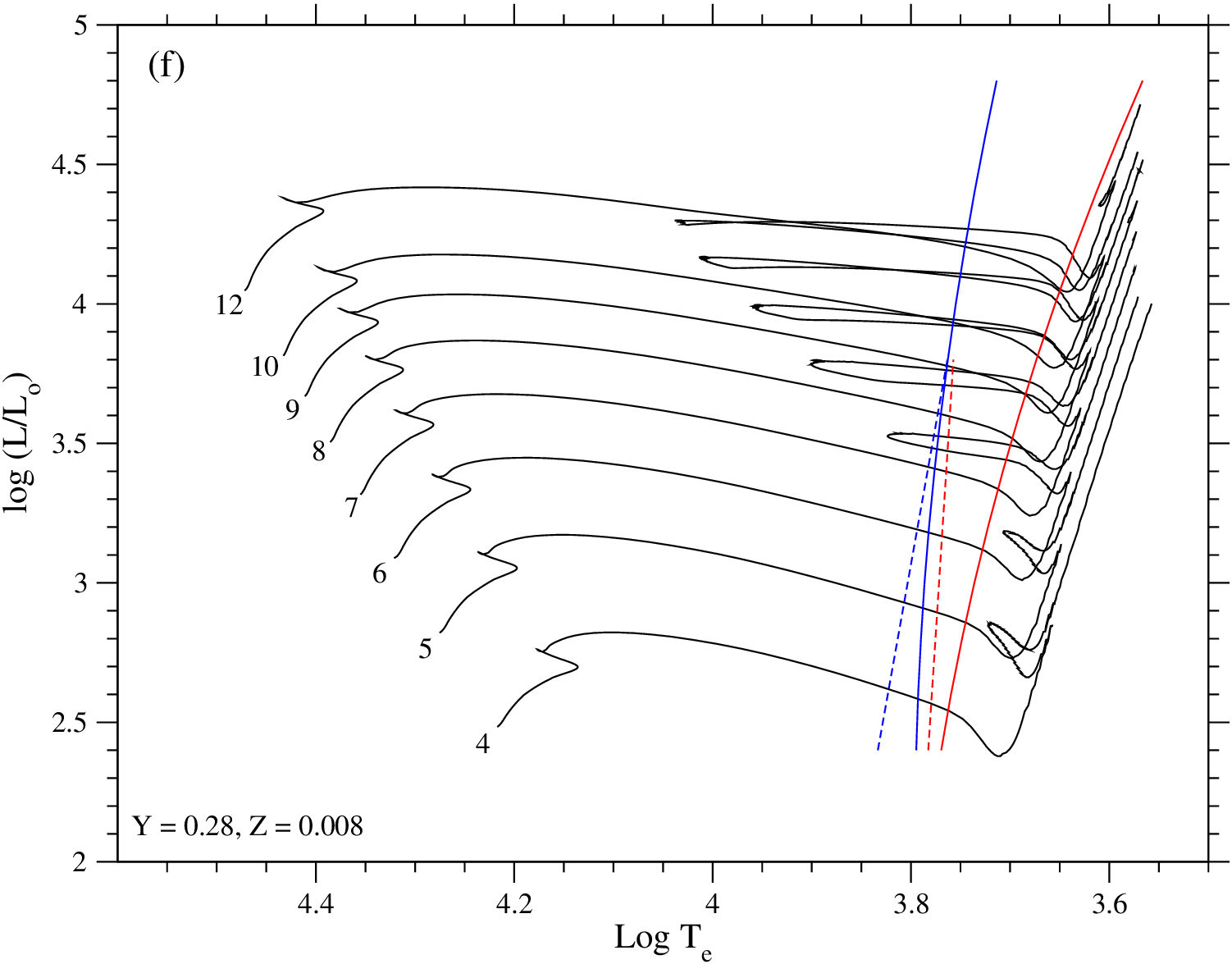}\vspace{2mm}
\caption{Panels (a) to (f): Evolutionary tracks, quadratic fits of the fundamental mode instability strip boundaries and
linear fits of the first overtone ones, for the cases I to VI of
Table~\ref{table:casitrattati}.}
\label{fig:set-tracce}
\end{figure*}

Inspection of Fig.~\ref{fig:set-tracce} suggests that the loop extension generally increases with increasing
  mass, but,  in some cases, it decreases for the 
5~M$_{\odot}$ model. As discussed by \citet{castchief00} this is 
because for masses around this value the first dredge-up (which is active for
lower masses) becomes inefficient and the star reaches the central He burning
phase with a reduced external He abundance (and thus with a reduced loop
extension) with respect to the lower masses.

For each model set we found the minimum (M$_{min}$) and maximum (M$_{max}$) mass which
cross the instability strip.  The M$_{max}$ and M$_{min}$ values are listed in Table~
\ref{table:mminmmax}, except for the cases including overshooting. The minimum mass in general increases with the metallicity and the helium content while
the maximum mass increases for a decrease of the
metallicity and an increase of $Y$.

\begin{table}
\caption{Maximum (M$_{max}$) and minimum (M$_{min}$) mass entering
the instability strip, for the various model sets.}
    \label{table:mminmmax}   
\centering                     
\begin{tabular}{l l l }       
\hline\hline
Case & M$_{min}$ (M$_{\odot}$) & M$_{max}$ (M$_{\odot}$)  \\  
\hline  
I  ($std$)       & 4 & 12  \\
II ($Z_{low}$)      & 3 & 14  \\
III ($Z_{high}$)      & 4 &  8 \\ 
IV ($Y_{high}$)       & 4 & 13  \\ 
\hline                           
\end{tabular}
\end{table}

To derive the $ML$ relation \citep[see e.g.][]{B00} we take into
account the three crossings of the instability strip: the first crossing
during the H burning and the second and third crossings during the He burning
phases.  The mean luminosity is obtained by averaging the luminosity behaviour
during the three subsequent crossings with the corresponding evolutionary
times.  Following this procedure, for each case from I to IV of
Table~\ref{table:casitrattati} we derived the corresponding $ML$ relation (the results are available on line in Table~\ref{table:MLrelations}).

\onltab{6}{
\begin{table}
\caption{$ML$ relations for the selected  cases; $\log (L/L_{\odot}) = a + b\; \log (M/M_{\odot})$.} \label{table:MLrelations}  
\centering 
\begin{tabular}{l l l l }       
\hline\hline
Case & $a$ & $b$ & $\sigma$\\  
\hline
I  ($std$)      & 1.000 $\; \pm \;$ 0.017 & 3.160  $\; \pm \;$ 0.019 & 0.009 \\
II ($Z_{low}$)     & 1.040 $\; \pm \;$ 0.010 & 3.110  $\; \pm \;$ 0.011 & 0.008 \\
III ($Z_{high}$)     & 0.736 $\; \pm \;$0.008& 3.455  $\; \pm \;$ 0.010 & 0.002 \\
IV ($Y_{high}$)      & 1.095 $\; \pm \;$ 0.043 & 3.074  $\; \pm \;$ 0.047 & 0.024 \\
\hline                           
\end{tabular}
\end{table}
}

Since one of the purposes of this work is to investigate the spread of the
$ML$ relation due to uncertainties on the chemical composition, we calculated
the global $ML$ relation by fitting together all the $ML$ data sets (cases I-IV of Table~\ref{table:casitrattati}).
As a result we find:
 
\begin{equation}\label{mltotale}
\log (L/L_{\odot}) = 1.015 (\pm 0.021) + 3.150 (\pm 0.024) \; \log (M/M_{\odot})
\end{equation}
with a standard deviation of $\sigma= 0.025$;
the errors include the effect of varying the chemical abundances among the different cases. 
The comparison with the $ML$ relations obtained for the individual cases separately (see Table~\ref{table:MLrelations} on line) suggest that the coefficients of the global $ML$ relation are consistent with the ones of the individual relations apart from  case  III ($Z_{high}$) that corresponds to a significantly steeper relation, with a fainter zero-point, than all the others. This occurrence produces an increase of the intrinsic dispersion when passing from the individual $ML$ relations to the global one.

To compare the present result with a previous published relation based on evolutionary and pulsation models including older physical inputs, we compared the present $ML$ relation, given by
Eq.~\ref{mltotale}, with the one by \citet{B00}, which
explicitly takes into account the dependence on the chemical composition, computed for the
values of $Y$ and $Z$ corresponding to our standard case: $Y$=0.26, $Z$=0.005.
We found that the $ML$ relation derived in this paper is flatter than \citet{B00} with the disagreement reaching about 0.1 dex at masses around 10 M$_{\odot}$. However for M$\le{\sim{8{\rm M}_{\odot}}}$ the two relations are consistent within the uncertainties.

\subsection{Period-Luminosity relations}\label{sec:PL}

For the cases I to IV of Table~\ref{table:casitrattati} we calculated the
$PLT_e$ relations for the fundamental and first overtone models; the tables with the
coefficients for these relations are available on line in Table~\ref{table:can-PLT}.

\onltab{7}{
\begin{table*}
\bigskip
\caption{Canonical $PLT_e$ relations for the selected  cases; $\log P = a + b\; \log(L/L_{\odot}) + c\; (\log T_{e})$.}      
\label{table:can-PLT}  
\centering 
\begin{tabular}{l l l l l}       
\hline\hline
Case & $a$ & $b$ & $c$ & $\sigma$\\  
\hline 
Fundamental - CANONICAL\\
\hline
I  ($std$)  & 10.725 $\; \pm \;$ 0.180 & 0.687  $\; \pm \;$  0.004  &-3.261  $\; \pm \;$0.046 & 0.013 \\
II ($Z_{low}$) & 10.475 $\; \pm \;$ 0.320 &0.696   $\; \pm \;$0.005   & -3.202 $\; \pm \;$0.082 & 0.016 \\
III  ($Z_{high}$)& 10.908 $\; \pm \;$ 0.173 & 0.677 $\; \pm \;$ 0.004  &-3.303  $\; \pm \;$0.044 &0.009  \\
IV  ($Y_{high}$) & 11.265 $\; \pm \;$0.270  & 0.681 $\; \pm \;$ 0.005  &-3.398  $\; \pm \;$0.069 &0.013  \\
\hline
First Overtone - CANONICAL\\
\hline
I  ($std$) & 10.827 $\; \pm \;$0.393   &0.633 $\; \pm \;$0.004   & -3.282 $\; \pm \;$0.101 &0.003  \\
II ($Z_{low}$)& 10.826 $\; \pm \;$0.187  &0.631  $\; \pm \;$0.002   & -3.279 $\; \pm \;$0.048 &0.002   \\
III  ($Z_{high}$)&11.033 $\; \pm \;$0.404  &0.630  $\; \pm \;$0.003   & -3.337 $\; \pm \;$0.105 &0.004  \\
IV  ($Y_{high}$) &10.897 $\; \pm \;$0.347  &0.633  $\; \pm \;$0.003   & -3.299 $\; \pm \;$0.090 &0.002  \\
\hline
\end{tabular}
\end{table*}
}

We derived the synthetic $PL$ relations by populating the predicted
instability strip according to an assumed mass distribution and
adopting, for each mass and chemical composition, the corresponding $ML$
relation derived on the basis of the corresponding set of evolutionary tracks
(see Subsec.\ref{sec:ML}).
More in detail, for each treated case (I to IV of Table~\ref{table:casitrattati}), 
we built a synthetic population by extracting masses according to a Salpeter 
initial mass function (IMF) ($\propto M^{-2.35}$). 
 The range of generated masses is given
by the values of M$_{max}$ and M$_{min}$ for the cases I to IV listed in
Table~\ref{table:mminmmax}. 
In agreement with the procedure adopted in previous papers \citep{cmm00, k98}, we fixed to $N$=1000  the total number of extractions falling into the
instability strip. To each synthetic star we assigned the luminosity derived from the
corresponding $ML$ relation.  
 Once the luminosity is fixed, from the quadratic fits of the fundamental edges,
 derived in Subsec.\ref{sec:strip}, we obtained the corresponding temperature of the 
FRE and the FBE. In the temperature range from the $T_e$ of the FBE (decreased by 50 K)
to the $T_e$ of the FRE (increased by 50 K), we then extracted randomly the
temperature $T_e$ of the synthetic star. 
For each extracted $T_e$ (in K) the period (in days) is obtained from the $P = f(L,M,T_e)$
relation, resulting from a linear regression  
through the fundamental models of each selected case. 
Then, by means of the static model atmospheres by \citet{bess}, for the explored chemical compositions, we converted the model intrinsic luminosities into the various photometric bands, obtaining synthetic distributions in period- magnitude planes. The $PL$ relations are derived by means of linear (or quadratic) regressions through these model distributions.\\

\begin{table}
\caption{Maximum (P$_{max}$) minimum (P$_{min}$) periods in days 
derived from the synthetic Cepheid distributions.}     
\label{table:pminpmax}   
\centering                     
\begin{tabular}{l l l }       
\hline\hline
Case & P$_{min}$ (days)  & P$_{max}$ (days)  \\  
\hline  
I  ($std$)       & 2.5  & 72.9 \\
II ($Z_{low}$)      & 1.4 & 87.6  \\
III ($Z_{high}$)      & 2.1 & 21.9 \\ 
IV ($Y_{high}$)       & 2.6 & 92.0  \\ 
\hline                        
\end{tabular}
\end{table}

\begin{table}
\bigskip
\caption{Theoretical linear period-luminosity relations in different
  photometric bands for fundamental pulsators in the standard and global cases.
}      
\label{table:relazioniPL}  
\centering 
\begin{tabular}{l l l l}       
\hline
\hline
Case & $a$ & $b$ & $\sigma$\\  
\hline 
&&$\overline{M_B}$&\\
\hline
$std$     & -1.04 $\pm$ 0.03 & -2.30 $\pm$ 0.03 & 0.32 \\
$all$     & -0.87 $\pm$ 0.01 & -2.43 $\pm$ 0.01 & 0.32 \\
$std$ $cut$ & -0.95 $\pm$ 0.03 & -2.42 $\pm$ 0.03 & 0.30 \\
$all$ $cut$ & -0.83 $\pm$ 0.01 & -2.49 $\pm$ 0.02 & 0.30 \\
\hline
\hline 
&&$\overline{M_V}$&\\
\hline
$std$     & -1.40 $\pm$ 0.02 & -2.69 $\pm$ 0.02 & 0.23 \\
$all$     & -1.28 $\pm$ 0.01 & -2.77 $\pm$ 0.01 & 0.23 \\
$std$ $cut$ & -1.34 $\pm$ 0.02 & -2.77 $\pm$ 0.02 & 0.22 \\
$all$ $cut$ & -1.25 $\pm$ 0.01 & -2.82 $\pm$ 0.01 & 0.22 \\
\hline
\hline 
&&$\overline{M_R}$&\\
\hline
$std$     & -1.65 $\pm$ 0.02 & -2.83 $\pm$ 0.02 & 0.20 \\
$all$     & -1.55 $\pm$ 0.01 & -2.91 $\pm$ 0.01 & 0.20 \\
$std$ $cut$ & -1.59 $\pm$ 0.02 & -2.91 $\pm$ 0.02 & 0.18 \\
$all$ $cut$ & -1.52 $\pm$ 0.01 & -2.95 $\pm$ 0.01 & 0.19 \\
\hline
\hline 
&&$\overline{M_I}$&\\
\hline
$std$     & -1.92 $\pm$ 0.01 & -2.96 $\pm$ 0.02 & 0.17 \\
$all$     & -1.84 $\pm$ 0.01 & -3.01 $\pm$ 0.01 & 0.17 \\
$std$ $cut$ & -1.87 $\pm$ 0.02 & -3.02 $\pm$ 0.02 & 0.15 \\
$all$ $cut$ & -1.82 $\pm$ 0.01 & -3.05 $\pm$ 0.01 & 0.16 \\
\hline
\hline 
&&$\overline{M_J}$&\\
\hline
$std$     & -2.238 $\pm$ 0.010 & -3.149 $\pm$ 0.011 & 0.12 \\
$all$     & -2.184 $\pm$ 0.005 & -3.184 $\pm$ 0.005 & 0.12 \\
$std$ $cut$ & -2.206 $\pm$ 0.011 & -3.191 $\pm$ 0.012 & 0.11 \\
$all$ $cut$ & -2.166 $\pm$ 0.005 & -3.210 $\pm$ 0.006 & 0.11 \\
\hline 
\hline 
&&$\overline{M_K}$&\\
\hline
$std$     & -2.499 $\pm$ 0.006 & -3.329 $\pm$ 0.006 & 0.07 \\
$all$     & -2.470 $\pm$ 0.003 & -3.343 $\pm$ 0.003 & 0.08 \\
$std$ $cut$ & -2.481 $\pm$ 0.007 & -3.353 $\pm$ 0.008 & 0.07 \\
$all$ $cut$ & -2.459 $\pm$ 0.003 & -3.359 $\pm$ 0.004 & 0.07 \\
\hline
\end{tabular}
\end{table}

\onltab{10}{
\begin{table}[ht]
\caption{Theoretical linear period-luminosity relations in different
  photometric bands for individual cases.}\label{table:relazioniPL-online}
\begin{center}
\begin{tabular}{llll}
  \hline
Case & $a$ & $b$ & $\sigma$ \\
  \hline
&&$\overline{M_B}$&\\
\hline
  $Z_{low}$ & -0.72$\; \pm \;$0.02 & -2.61$\; \pm \;$0.02 & 0.27 \\
  $Z_{high}$ & -1.00$\; \pm \;$ 0.03 & -2.18$\; \pm \;$0.04 & 0.29 \\
  $Y_{high}$ & -1.03$\; \pm \;$0.03 & -2.28$\; \pm \;$0.03 & 0.37 \\
\hline
\hline
&&$\overline{M_V}$&\\
\hline
  $Z_{low}$ & -1.16$\; \pm \;$0.01 & -2.90$\; \pm \;$0.01 & 0.20 \\
  $Z_{high}$ & -1.40$\; \pm \;$0.02 & -2.61$\; \pm \;$0.03 & 0.21 \\
  $Y_{high}$ & -1.39$\; \pm \;$0.02 & -2.68$\; \pm \;$0.02 & 0.27 \\
\hline
\hline
&&$\overline{M_R}$&\\
\hline
  $Z_{low}$ & -1.44$\; \pm \;$0.01 & -3.00$\; \pm \;$0.01 & 0.16 \\
  $Z_{high}$ & -1.66$\; \pm \;$0.02 & -2.77$\; \pm \;$0.02 & 0.18 \\
  $Y_{high}$ & -1.63$\; \pm \;$0.02 & -2.82$\; \pm \;$0.02 & 0.23 \\
\hline
\hline 
&&$\overline{M_I}$&\\
\hline
  $Z_{low}$ & -1.75$\; \pm \;$0.01 & -3.10$\; \pm \;$0.01 & 0.14 \\
  $Z_{high}$ & -1.93$\; \pm \;$0.02 & -2.90$\; \pm \;$0.02 & 0.15 \\
  $Y_{high}$ & -1.90$\; \pm \;$0.02 & -2.95$\; \pm \;$0.02 & 0.20 \\
\hline
\hline 
&&$\overline{M_J}$&\\
\hline
  $Z_{low}$ & -2.117$\; \pm \;$0.006 & -3.240$\; \pm \;$0.007 & 0.10 \\
  $Z_{high}$ & -2.247$\; \pm \;$0.011 & -3.119$\; \pm \;$0.014 & 0.11 \\
  $Y_{high}$ & -2.217$\; \pm \;$0.013 & -3.140$\; \pm \;$0.013 & 0.14 \\
\hline 
\hline 
&&$\overline{M_K}$&\\
\hline
  $Z_{low}$ & -2.430$\; \pm \;$0.003 & -3.373$\; \pm \;$0.004 & 0.06 \\
  $Z_{high}$ & -2.507$\; \pm \;$0.007 & -3.317$\; \pm \;$0.009 & 0.07 \\
  $Y_{high}$ & -2.477$\; \pm \;$0.008 & -3.320$\; \pm \;$0.008 & 0.09 \\
   \hline
\end{tabular}
\end{center}
\end{table}
}
The maximum and minimum periods corresponding to the obtained synthetic distributions are reported in Table \ref{table:pminpmax}.
In Tables \ref{table:relazioniPL} and \ref{table:relazioniPLquadratiche} we report the linear ($\overline{M_\lambda} = a\; +\; b\; \log P$)
and quadratic ($\overline{M_\lambda} = a\; +\; b \; \log P \; +\;
c \; (\log P)^2$) period-luminosity relations for fundamental
pulsators in the different photometric bands. In these Tables we listed
the fit coefficients for case I of Table~\ref{table:casitrattati}
({\em std}) and the global fit of cases I to IV taken together ({\em
  all}), as well as the results for the $PL$ relations obtained by
rejecting periods longer than $\log P \geq 1.5$ ({\em cut}).
The case {\em all} provides an  estimate of the effect 
of the uncertainties on the chemical abundances on the predicted $PL$ relations.
The linear and quadratic $PL_{\lambda}$ relations, computed for the single cases
II, III and IV of Table~\ref{table:casitrattati}, are available on
line in Tables \ref{table:relazioniPL-online} and
\ref{table:relazioniPL-quad-online}. \\

Passing from the individual cases to the global $PL$ relations, the coefficients and the intrinsic dispersion do not vary significantly because the effects of these variations in the abundances are smaller than the effect on the intrinsic scatter due to the finite width of the instability strip, especially in the optical filters. Moreover, inspection of Table
\ref{table:relazioniPL} suggests that the $PL$ relations become more linear, less dependent on
 the chemical composition and with a smaller intrinsic dispersion
 passing from the optical to the near-infrared filters,
 confirming previous observational \citep{mf91} and theoretical
 \citep{B99,cmm00} results. 
The $PL$ relations obtained for $\log P < 1.5$
are derived because, as shown in \citet{cmm00}, removing the longest periods,
the period-magnitude distribution is better reproduced by a linear relation
also in the optical bands. Moreover, many observed Cepheid samples cover this
period range. \\

\begin{figure}
\bigskip
\centering
\includegraphics[width=6.5cm]{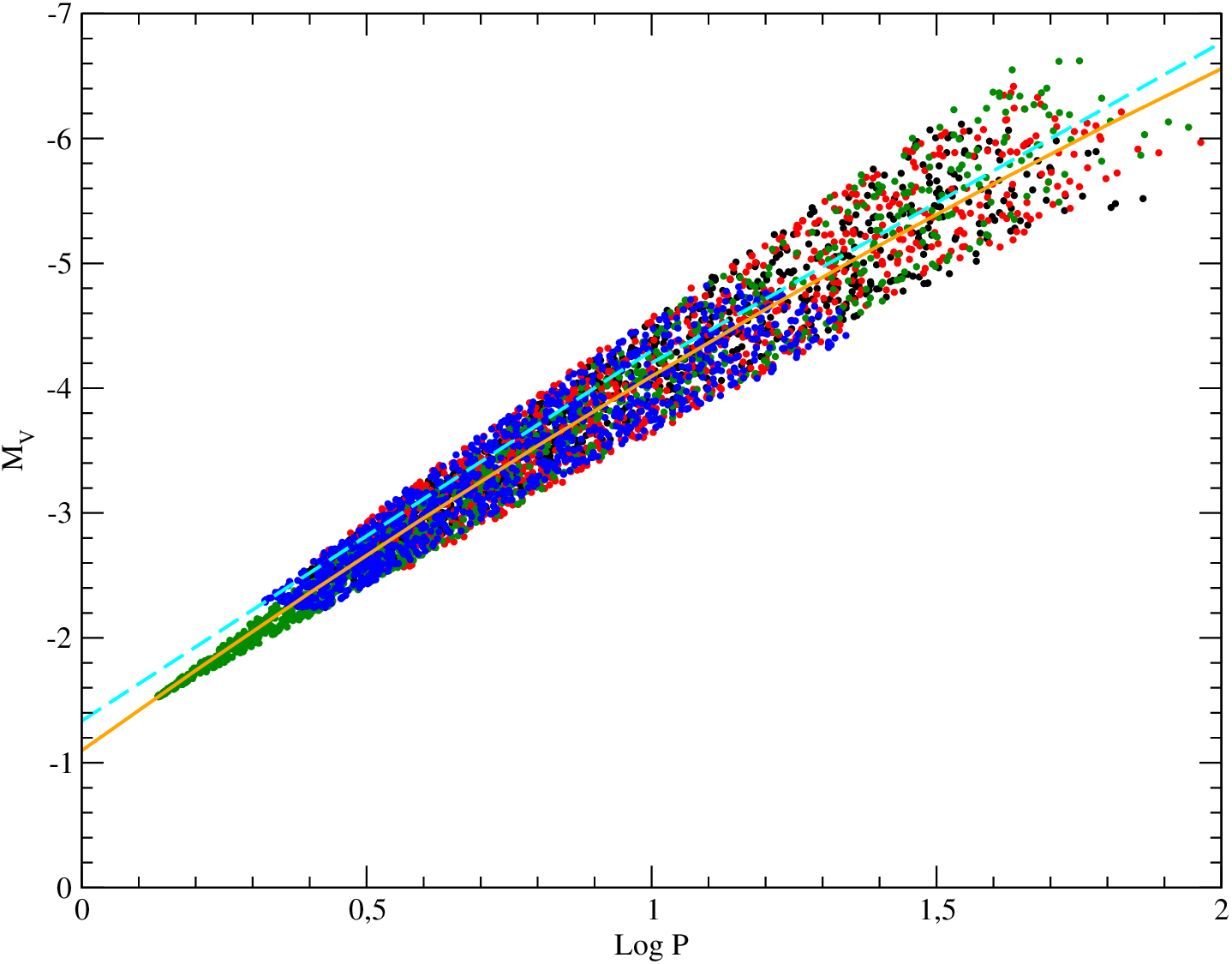}\hspace{3.5mm}
\includegraphics[width=6.5cm]{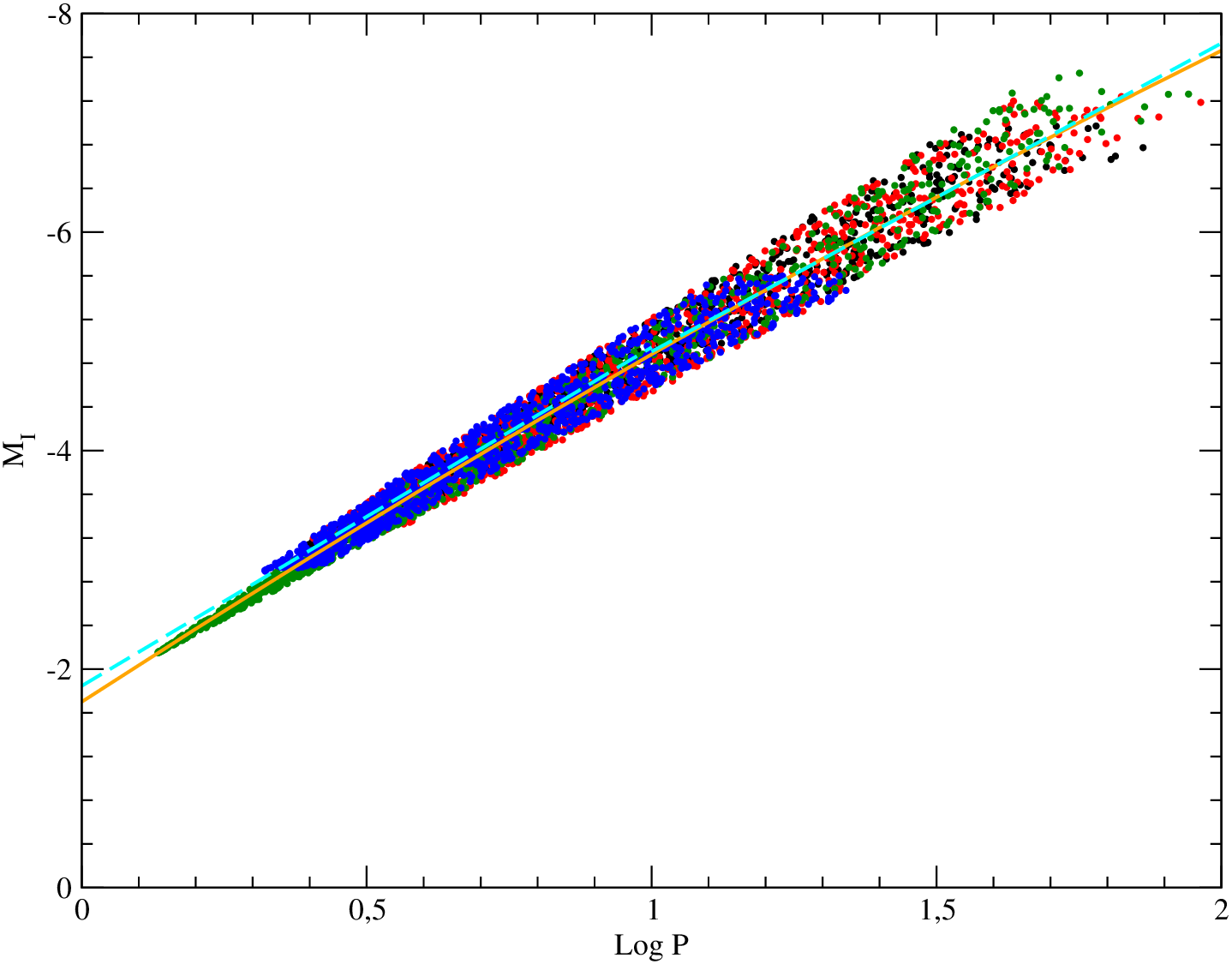}\vspace{5mm}\\
\caption{Synthetic period-luminosity relations in the V (upper panel) and I
  (lower panel) Johnson-Cousins photometric bands with the quadratic fits (solid lines)
corresponding to the case {\em all} of Table \ref{table:relazioniPLquadratiche}. Symbols: std case (case I, black dots); $Z_{low}$ (case II, green dots); $Z_{high}$ (case III, blue dots); $Y_{high}$ (case IV, red dots).}
\label{fig:PLall-nocut}
\end{figure}

\begin{figure}
\bigskip
\centering
\includegraphics[width=6.5cm]{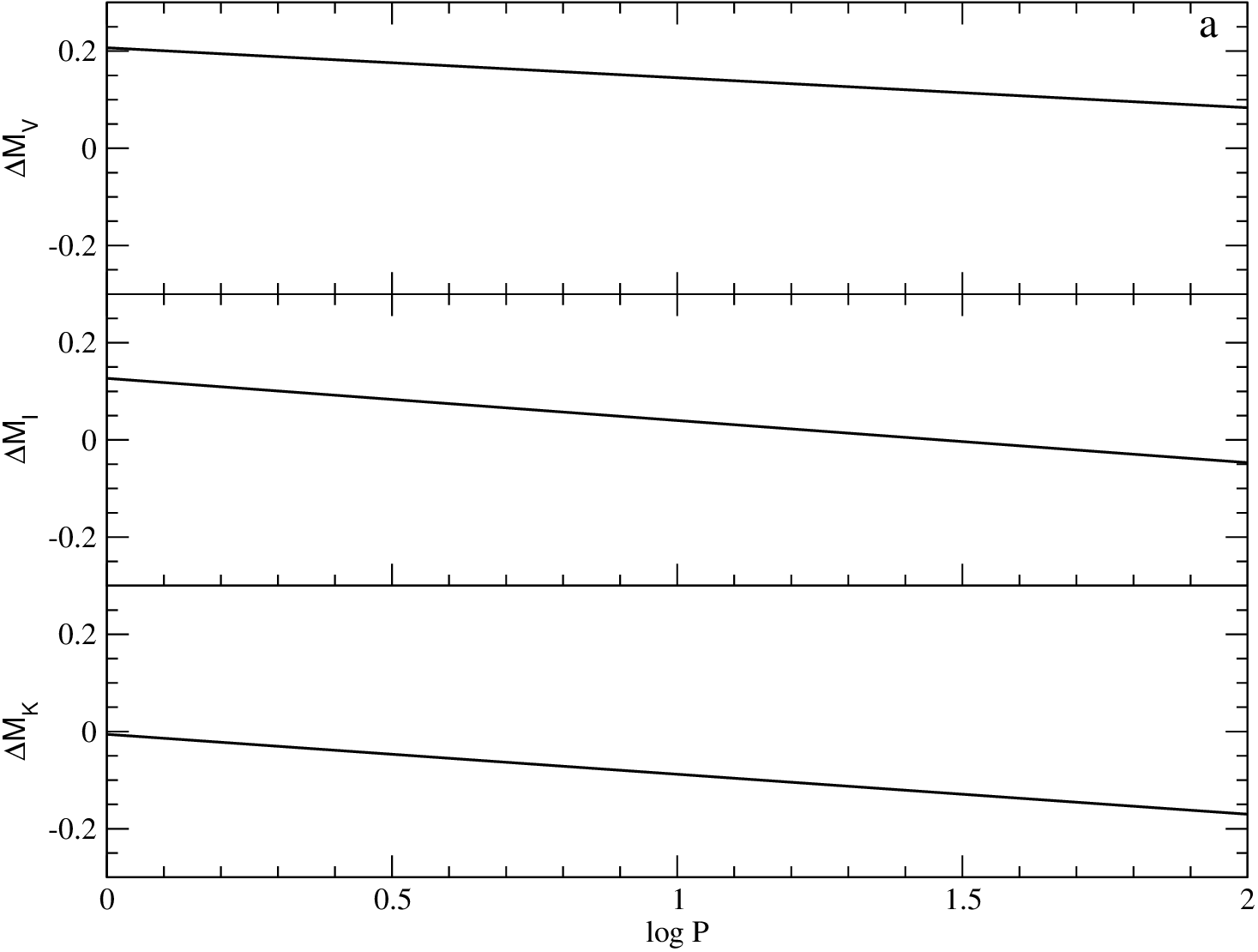}\hspace{3.5mm}
\includegraphics[width=6.5cm]{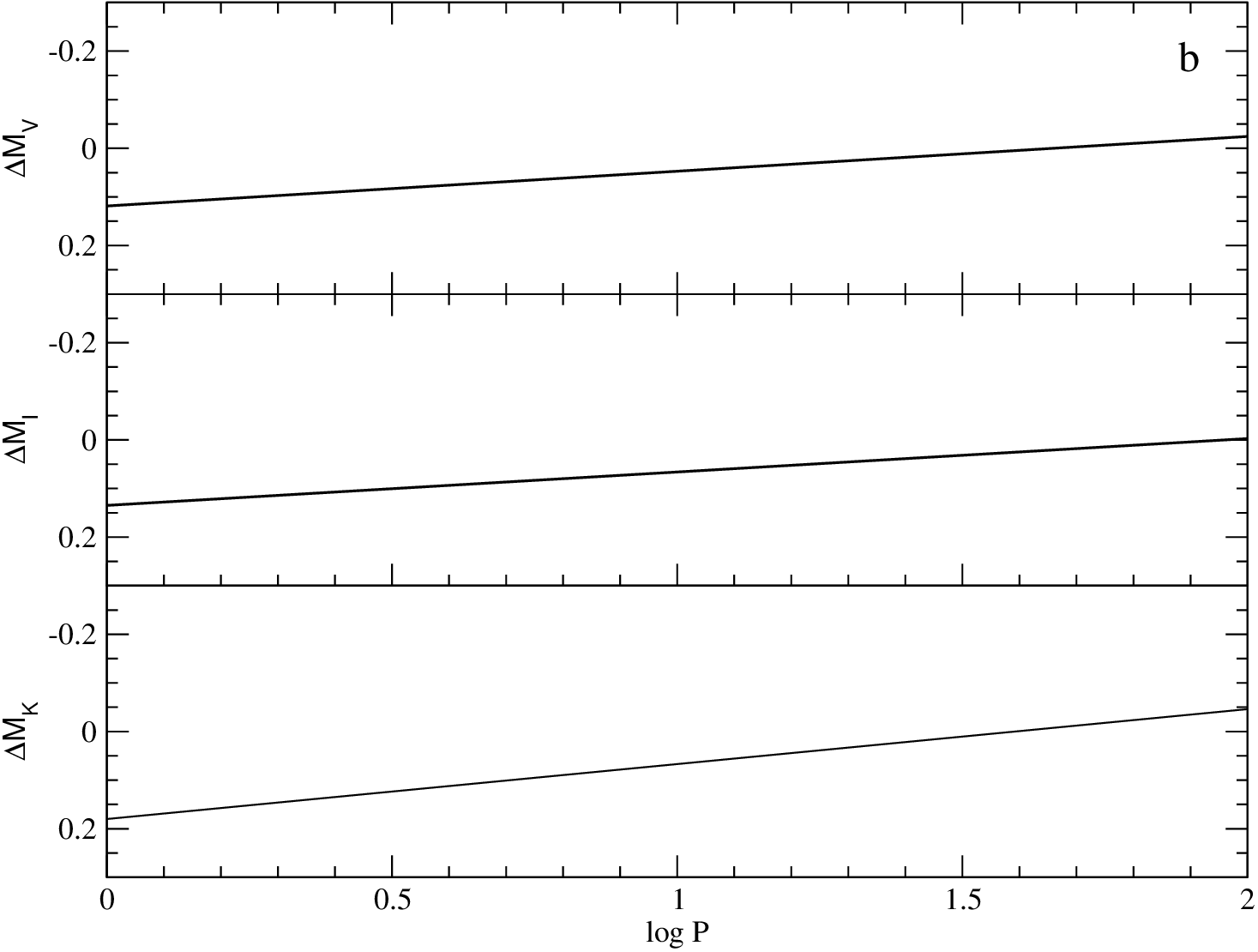}
\caption{a) Residuals of the present linear PL relations for 
  $\log P < 1.5$  in the V, I bands and over the whole period range for the K band with respect to the empirical relations; b) Differences between the present linear solutions with $\log P < 1.5$
and the theoretical ones by \citet{cmm00}.}
\label{fig:PL-VIK-confronto}
\end{figure}

The deviation from linearity of Cepheid $PL$ relations is a debated
issue \citep{sandage04,k04,n05,kn06} and for this reason we
also computed quadratic $PL$ relations.  
In particular, \citet{n05} performed a statistical investigation of
the LMC classical Cepheid sample obtained from the MACHO database and found that the
observed behaviour in period-magnitude diagrams is best reproduced by
two linear relations, with a break at 10 days. Nonlinear pulsation models \citep[see also][]{cmm00,f02,M05}
suggest a quadratic form of $PL$ relations, in particular in the optical
bands (see Table \ref{table:relazioniPLquadratiche}).  
However the effect of this nonlinearity on the calibration of
the extragalactic distance scale is very small \citep[see][]{k07}. 
In Fig.~\ref{fig:PLall-nocut} we plotted the predicted synthetic $PL$ relations in the V (upper panel) and I
  (lower panel) Johnson-Cousins photometric bands with the quadratic fits superimposed (solid lines) for cases I to IV taken together (case {\em all} in Table \ref{table:relazioniPLquadratiche}).
As a comparison  the two linear regressions (for $\log 
{P}<1.0$ and $\log{P} \ge 1.0$, dashed lines)
obtained observationally by \citet{sandage04} are also
shown. 
 In panel a of Fig.~\ref{fig:PL-VIK-confronto} we show the residuals of the present linear PL relations for 
  $\log P < 1.5$ (case {\em all cut} of Table \ref{table:relazioniPL})  in the V, I bands and over the whole period range for the K band with respect to the empirical relations by \citet{free01} (for V and I) and \citet{Perss} (for the K band). According to this plot, we find differences smaller than 0.2 mag (in absolute value) in all the bands. Panel b of Fig.~\ref{fig:PL-VIK-confronto} shows the differences
between the present linear solutions with $\log P < 1.5$
and the theoretical $PL$ linear relations  
  with $\log P < 1.5$ by \citet{cmm00}. Again we find differences within 0.2 mag with an opposite trend as a function of the period with respect to the previous comparison, thus suggesting that the the magnitudes predicted by the new PL relations obtained in this paper are somewhat intermediate between the ones based on empirical relations and the theoretical relations by \citet{cmm00}.

As far the physical assumptions of our pulsation models are concerned, we have already shown that changes in the adopted equation of state do not affect significantly the pulsation scenario \citep{pe03}. On the other hand, variations of the mixing length parameters used to close the nonlinear equation system \citep[see][]{BS94} have been found to produce negligible effects on the predicted Cepheid relations \citep[see][for details]{f07}.

\begin{table}
\bigskip
\caption{Theoretical quadratic period-luminosity relations for fundamental
pulsators in the standard and global cases.}
\label{table:relazioniPLquadratiche}  
\centering 
\begin{tabular}{l l l l l}       
\hline
\hline
Case & $a$ & $b$ & $c$ & $\sigma$\\  
\hline 
&&$\overline{M_B}$&\\
\hline
$std$ & -0.43 $\pm$ 0.08 & -3.69 $\pm$ 0.17 & 0.69 $\pm$ 0.08 & 0.31 \\
$all$ & -0.63 $\pm$ 0.02 & -3.07 $\pm$ 0.06 & 0.35 $\pm$ 0.03 & 0.32 \\
\hline
\hline 
&&$\overline{M_V}$&\\
\hline
$std$ & -0.98 $\pm$ 0.06 & -3.66 $\pm$ 0.12 & 0.48 $\pm$ 0.06 & 0.22 \\
$all$ & -1.10 $\pm$ 0.02 & -3.27 $\pm$ 0.04 & 0.27 $\pm$ 0.02 & 0.23 \\
\hline
\hline
&&$\overline{M_R}$&\\
\hline
$std$ & -1.29 $\pm$ 0.05 & -3.65 $\pm$ 0.10 & 0.41 $\pm$ 0.05 & 0.19 \\
$all$ & -1.39 $\pm$ 0.02 & -3.33 $\pm$ 0.04 & 0.23 $\pm$ 0.02 & 0.19 \\
\hline
\hline 
&&$\overline{M_I}$&\\
\hline
$std$ & -1.61 $\pm$0.04 & -3.66 $\pm$ 0.09 & 0.35 $\pm$ 0.04 & 0.16 \\
$all$ & -1.70 $\pm$0.01 & -3.38 $\pm$ 0.03 & 0.20 $\pm$ 0.02 & 0.16 \\
\hline
\hline 
&&$\overline{M_J}$&\\
\hline
$std$ & -2.03 $\pm$ 0.03 & -3.63 $\pm$ 0.06 & 0.24 $\pm$ 0.03 & 0.12 \\
$all$ & -2.08 $\pm$ 0.01 & -3.45 $\pm$ 0.02 & 0.14 $\pm$ 0.01 & 0.12 \\
\hline
\hline
&&$\overline{M_K}$&\\
\hline
$std$ & -2.38 $\pm$ 0.02 & -3.61 $\pm$ 0.04 & 0.14 $\pm$ 0.02 & 0.07 \\
$all$ & -2.41 $\pm$ 0.01 & -3.50 $\pm$ 0.01 & 0.09 $\pm$ 0.01 & 0.07 \\
\hline
\end{tabular}
\end{table}

\onltab{12}{
\begin{table}[ht]
\caption{Theoretical quadratic period-luminosity relations for fundamental
pulsators in the various individual cases.}\label{table:relazioniPL-quad-online}
\begin{center}
\begin{tabular}{lllll}
 \hline
\hline
Case & $a$ & $b$ & $c$ & $\sigma$\\  
\hline 
&&$\overline{M_B}$&\\
\hline
  $Z_{low}$ & -0.62$\; \pm \;$0.03 & -2.92$\; \pm \;$0.08 & 0.18$\; \pm \;$0.04 & 0.27 \\
  $Z_{high}$ & -0.25$\; \pm \;$0.08 & -4.32$\; \pm \;$0.22 & 1.35$\; \pm \;$0.14 & 0.27 \\
  $Y_{high}$ & -0.74$\; \pm \;$0.10 & -2.92$\; \pm \;$0.20 & 0.30$\; \pm \;$0.09 & 0.36 \\
\hline
\hline 
&&$\overline{M_V}$&\\
\hline
  $Z_{low}$ & -1.10$\; \pm \;$0.02 & -3.08$\; \pm \;$0.06 & 0.11$\; \pm \;$0.03 & 0.20 \\
  $Z_{high}$ & -0.87$\; \pm \;$0.06 & -4.12$\; \pm \;$0.16 & 0.96$\; \pm \;$0.10 & 0.20 \\
  $Y_{high}$ & -1.19$\; \pm \;$0.07 & -3.09$\; \pm \;$0.14 & 0.20$\; \pm \;$0.07 & 0.27 \\
\hline
\hline
&&$\overline{M_R}$&\\
\hline
  $Z_{low}$ & -1.40$\; \pm \;$0.02 & -3.15$\; \pm \;$0.05 & 0.08$\; \pm \;$0.03 & 0.16 \\
  $Z_{high}$ & -1.21$\; \pm \;$0.05 & -4.05$\; \pm \;$0.14 & 0.81$\; \pm \;$0.09 & 0.17 \\
  $Y_{high}$ & -1.47$\; \pm \;$0.06 & -3.17$\; \pm \;$0.12 & 0.16$\; \pm \;$0.06 & 0.23 \\
\hline
\hline 
&&$\overline{M_I}$&\\
\hline
  $Z_{low}$ & -1.71$\; \pm \;$0.02 & -3.22$\; \pm \;$0.04 & 0.07$\; \pm \;$0.02 & 0.14 \\
  $Z_{high}$ & -1.54$\; \pm \;$0.04 & -4.01$\; \pm \;$0.12 & 0.70$\; \pm \;$0.07 & 0.14 \\
  $Y_{high}$ & -1.76$\; \pm \;$0.05 & -3.25$\; \pm \;$0.10 & 0.14$\; \pm \;$0.05 & 0.19 \\
 \hline
\hline 
&&$\overline{M_J}$&\\
\hline
  $Z_{low}$ & -2.09$\; \pm \;$0.01 & -3.32$\; \pm \;$0.03 & 0.04$\; \pm \;$0.02 & 0.10 \\
  $Z_{high}$ & -1.97$\; \pm \;$0.03 & -3.90$\; \pm \;$0.08 & 0.50$\; \pm \;$0.05 & 0.10 \\
  $Y_{high}$ & -2.13$\; \pm \;$0.04 & -3.34$\; \pm \;$0.08 & 0.09$\; \pm \;$0.04 & 0.14 \\
\hline
\hline
&&$\overline{M_K}$&\\
\hline
  $Z_{low}$ & -2.42$\; \pm \;$0.01 & -3.41$\; \pm \;$0.02 & 0.02$\; \pm \;$0.01 & 0.06 \\
  $Z_{high}$ & -2.34$\; \pm \;$0.02 & -3.80$\; \pm \;$0.05 & 0.31$\; \pm \;$0.03 & 0.07 \\
  $Y_{high}$ & -2.43$\; \pm \;$0.02 & -3.43$\; \pm \;$0.05 & 0.05$\; \pm \;$0.02 & 0.09 \\
   \hline
\end{tabular}
\end{center}
\end{table}
}

\section{Noncanonical models}
\subsection{Overshooting} \label{oversh}

The extension of the
mixing into a region that is stable following the Schwarzschild criterion is generally called
``overshooting'' \citep[see e.g.][and references therein]{Cord02}.  An additional
extension of the canonical convective regions could also be due to rotationally
induced mixing \citep[see e.g.][]{Mey00}; both phenomena
can  be modeled, at least as far as the extension of the mixing zones is concerned, with the same
``overshooting'' formalism.  A complete understanding of these extra-mixing mechanisms is still lacking in the literature due to the
complexity of the involved phenomena, that also prevents an analytical
approach \citep[see e.g.][]{Stra05, Dem07}. 
Three-dimensional hydrodynamical calculations are still under development \citep[][]{Eggle07, Mea07}.
Thus a semi-empirical approach
is currently used  by defining the extension of any additional mixing region, starting from the border of the canonical
convective zones,  in terms of the pressure scale height $H_{\rm
  p}$: ${l}_{\rm ov}=\beta{H}_{\rm p}$.  The overshooting efficiency is generally
estimated through the comparison between theory and observations for different
relevant quantities, and is still widely debated
point \citep[see e.g.][and references therein]{Cla07, broc03, Barm02, Chio92}.

We calculated additional sets of
evolutionary models including overshooting during the central H burning phase
(Main Sequence, MS) with  ${l}_{\rm ov}=0.1 H_p$ and ${l}_{\rm ov}=0.25 H_p$,
within the range suggested in the recent literature \citep[see e.g.][]{Gir00}.
 During the central He burning phase our code implements the
classical semi-convection criterion \citep[see e.g.][]{Cast85} at the
border of the Schwarzschild convective core (which is driven by mechanical
overshooting at the convective core boundary), while breathing pulses \citep[see e.g.][]{Cast85,
Dor93} are suppressed following the
procedure described by  \citet{Cap89}.  In the present models the
overshooting region is fully homogenized but the
temperature gradient is kept at its radiative value.

The inclusion of overshooting during central H burning in standard models
leads to a higher luminosity and longer lifetime during the MS phase, larger
He cores and luminosities and reduced lifetimes during the following central
He burning phase.  This reflects in the adoption of a different
mass-luminosity relation for Cepheid pulsational models.  The adopted
luminosity levels for noncanonical models in this paper are taken as brighter
than the canonical ones by 0.25 dex , following the prescriptions by
\citet{cwc93}.  We checked that it is true for current evolutionary models
too.

When the overshooting phenomenon is included, the blue loops in the color-magnitude
diagram are less extended than in the absence of overshooting in such a way
that for some chemical compositions intermediate-mass models do not cross the
classical Cepheid instability strip. To avoid this problem, some authors 
include an ``ad hoc'' undershooting at the lower boundary of convective
envelopes \citep[see][]{Gir00}, even if there is no clear observational
evidence for this mechanism  \citep[see e.g.][]{Ren94, Rie03}.

Relevant pulsation properties, such as the morphology and the amplitude of
light and radial velocity curves, are affected by the assumed $ML$ relation
\citep{bcm02,n08}, and, in turn, by the amount of overshooting accounted for
in the models \citep[][and references therein]{c05,Kel08}. Independent
theoretical and observational evidence suggests that the Cepheid pulsation
masses are lower than the canonical evolutionary values but the amount of
this discrepancy is not firmly established even if several authors find a
value of the order of 10-15 $\%$.  \citep[e.g.][and references
therein]{b01,bbk01,c05,n08}.

To quantify the effect of overshooting, we computed
models with standard chemical composition ($Z=0.005$,
$Y=0.26$) for two selected values of the overshooting efficiency
($\beta$=0.1 and $\beta$=0.25, models XI and XIII of
  Table~\ref{table:casitrattati}). 
However, due to the
reduced loop extension, some intermediate mass models
with $\beta$=0.25 do not cross the instability strip, in
disagreement with observations of LMC Cepheids in this region. To avoid this problem
without including overshooting for the convective envelopes, taking into
account the present uncertainty in the original helium abundance, we
calculated overshooting models for $Z=0.005$, $Y=0.28$ (model XII of
Table~\ref{table:casitrattati}) which, as already
discussed, show a more extended loop than the standard models. In
this way, even models with $\beta$=0.25 populate the instability strip.

In Table~\ref{table:fitnoncanonico} the coefficients of the quadratic fits of
the FRE and FBE for models with $l_{\rm ov}=0.25 H_{\rm p}$ and standard
chemical composition are reported.

\begin{table}
\bigskip
\caption{Quadratic analytical relations for the 
  fundamental edges, in the non canonical standard case.}       
\label{table:fitnoncanonico}  
\centering 
\begin{tabular}{l l l l}
\hline\hline
 $a$ & $b$ & $c$ & $\sigma$\\
\hline
&Blue Edge&&\\
 \hline 
 3.904 $\; \pm \;$  0.046 &-0.034  $\; \pm \;$ 0.027  & -0.0001 $\; \pm \;$ 0.0038 & 0.002 \\
\hline    
&Red Edge&&\\
 \hline 
3.816 $\; \pm \;$  0.078  & 0.008 $\; \pm \;$ 0.046  & -0.0124 $\; \pm \;$ 0.0065  & 0.004 \\
\hline   
\end{tabular}
\end{table}

\begin{figure}
\bigskip
\centering
\includegraphics[width=6.5cm]{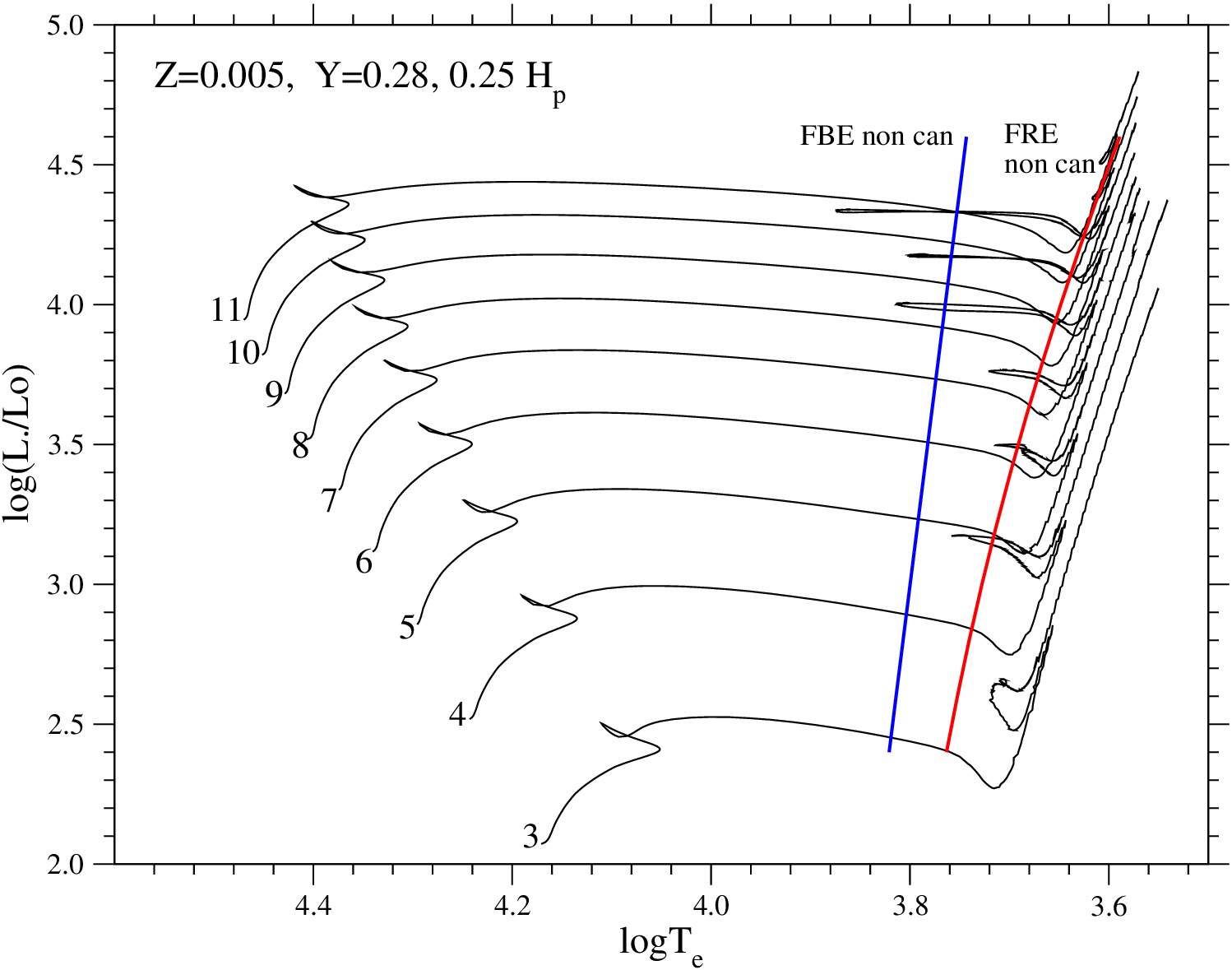}\vspace{4mm}
\includegraphics[width=6.5cm]{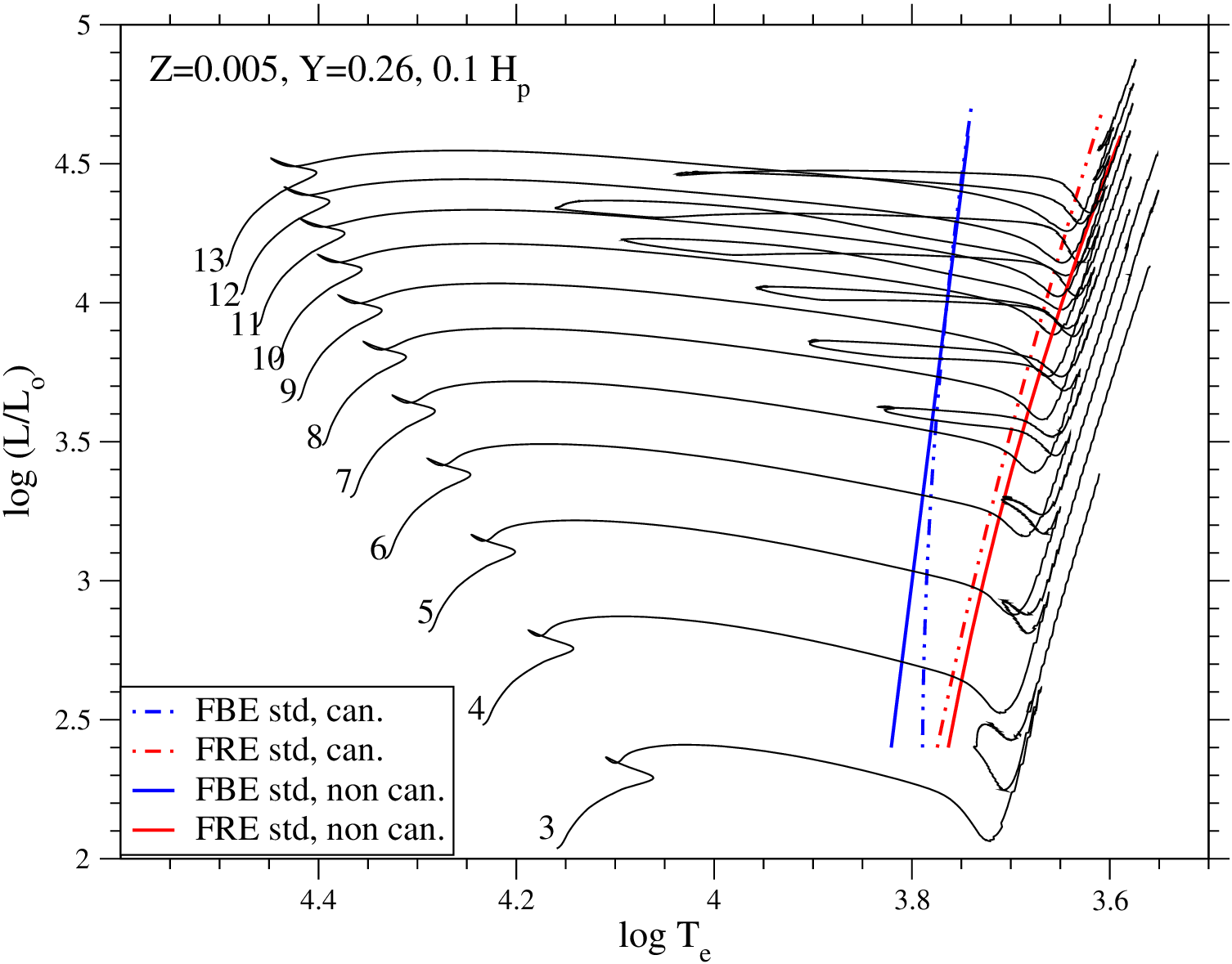}\vspace{5mm}\\
\caption{Evolutionary tracks with $Y$=0.28, $Z$=0.005, $l_{\rm ov}=0.25 H_{\rm
    p}$ and the quadratic fit of the noncanonical instability strip boundaries
(upper panel); evolutionary tracks for $Y$=0.26, $Z$=0.005, $l_{\rm ov}=0.1 H_{\rm p}$ and the quadratic fit
of the canonical (dot-dashed lines) and noncanonical (solid lines) instability
strip boundaries for standard chemical composition (lower panel).}\label{fig:os025}
\end{figure}

The upper panel of Fig.~\ref{fig:os025} shows the evolutionary tracks with
$l_{\rm ov}=0.25 H_{\rm p}$, $Z$=0.005, $Y$=0.28 and the quadratic fit of the
non canonical strip, while the lower panel of the same figure shows the
evolutionary tracks with $l_{\rm ov}=0.1 H_{\rm p}$, $Z$=0.005, $Y$=0.26 and
the quadratic fits of the canonical (dot-dashed lines) and non canonical
(solid lines) instability strip boundaries calculated for standard chemical
composition.  We notice that the case with $\beta$=0.1 is not significantly
different from the canonical one and for this reason in the following we will
concentrate on the case with $\beta$=0.25.

\begin{figure}
\bigskip
\centering
\includegraphics[width=8.5cm]{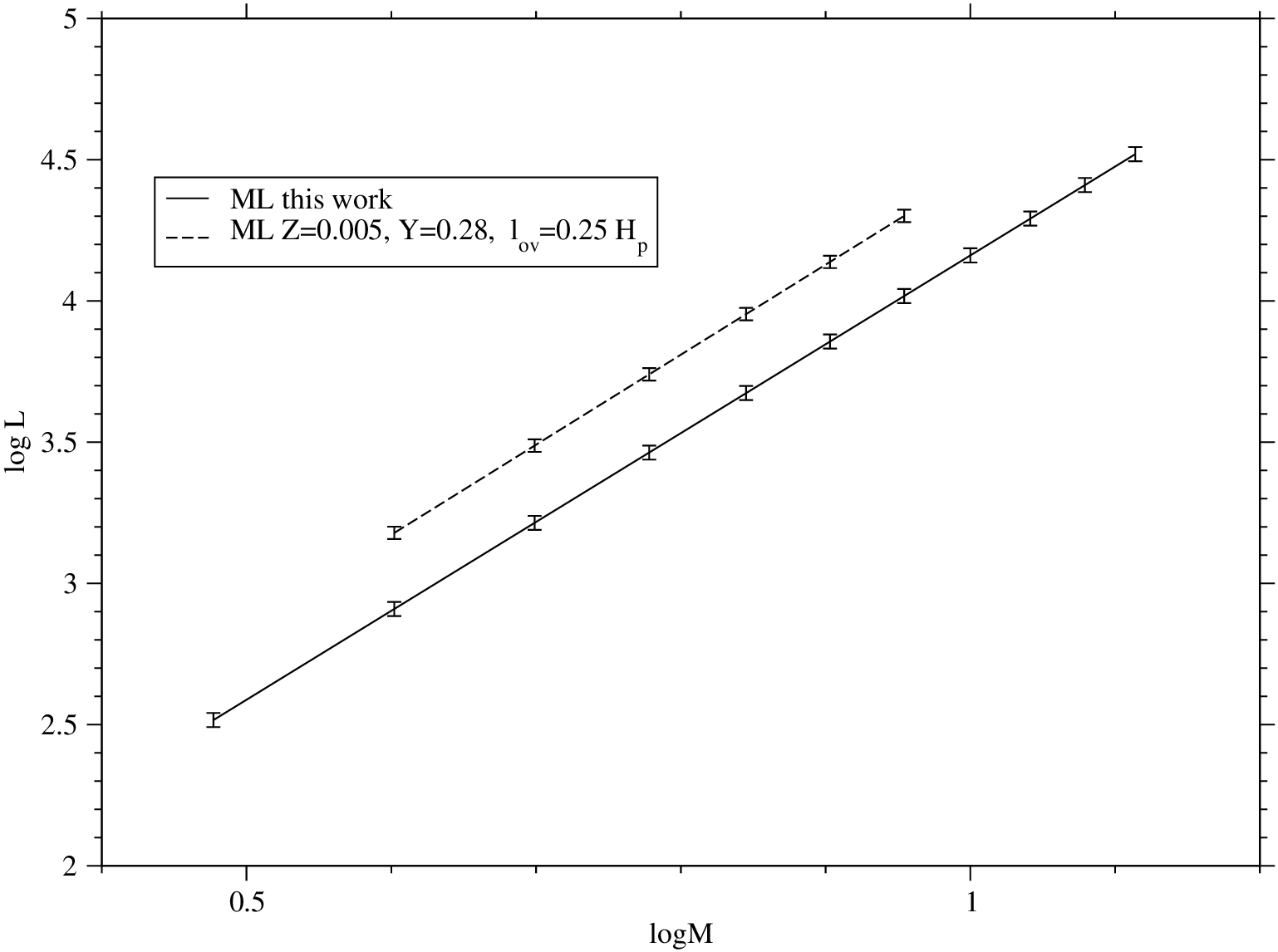}\hspace{3mm}
\caption{Comparison between the mass-luminosity relation from Eq.~\ref{ml-os025hp} ($Y$=0.28, $Z$=0.005, $l_{\rm ov}=0.25 H_{\rm p}$)
and the global $ML$ relation from Eq.~\ref{mltotale}}\label{fig:MLos025}.
\end{figure}

The $ML$ obtained with the set of tracks with  $l_{\rm ov}=0.25 H_{\rm p}$, $Z$=0.005, $Y$=0.28 combined with the noncanonical strip of
Fig.~\ref{fig:os025}, upper panel, is the following:
\begin{equation}\label{ml-os025hp}
\log (L/L_{\odot}) = 1.260 (\pm 0.061) +  3.187 (\pm 0.076) \; \log (M/M_{\odot})
\end{equation}
 with $\sigma= 0.022$.
 As expected this $ML$ relation has a brighter zero-point than the canonical one (see Eq.~\ref{mltotale}) by about 0.25 dex, and the same slope within the errors. This difference can be also noted from the comparison shown in Fig.~\ref{fig:MLos025}.

For the chemical composition of case I, in Table~\ref{table:casitrattati} we calculated the
$PLT_e$ relations for noncanonical fundamental and first overtone models; the tables with the
coefficients for these relations are available on line in Table~\ref{table:noncan-PLT}.

\onltab{14}{
\begin{table*}
\bigskip
\caption{Non canonical $PLT_e$ relations: $\log P = a + b\;
  \log(L/L_{\odot}) + c\; (\log T_{e})$.}      
\label{table:noncan-PLT}  
\centering 
\begin{tabular}{l l l l l}       
\hline\hline
Case & $a$ & $b$ & $c$ & $\sigma$\\  
 \hline 
Fundamental - Non canonical \\
\hline
I  ($std$)       & 10.290 $\; \pm \;$ 0.290  & 0.701 $\; \pm \;$ 0.005  & -3.144 $\; \pm \;$ 0.075 & 0.018  \\
\hline
First Overtone - Non canonical\\
\hline
I  ($std$)       & 10.840 $\; \pm \;$ 0.130  & 0.639 $\; \pm \;$ 0.002  & -3.278 $\; \pm \;$ 0.032 & 0.001  \\
\hline
\end{tabular}
\end{table*}
}

Adopting the same procedure as for canonical models we derived the multi-band period-luminosity relations for $\beta=0.25$, $Z$=0.005, $Y$=0.28. The coefficients of the linear $PL$ in the V, I, K bands  are reported in Table~\ref{table:relazioniPLlineari-OS}, whereas the linear and quadratic $PL$ relations in the other photometric bands are available on line, in Tables \ref{table:noncan-lin-PL} and \ref{table:noncan-quad-PL}.

\onltab{15}{
\begin{table}[ht]
\begin{center}
\caption{Theoretical linear period-luminosity relations for the fundamental noncanonical pulsators.}\label{table:noncan-lin-PL} 
\begin{tabular}{llll}
  \hline
Case & $a$ & $b$ & $\sigma$ \\
  \hline
&&$\overline{M_B}$&\\
\hline
$ncOS$ & -1.40$\; \pm \;$0.05 & -1.81$\; \pm \;$0.04 & 0.41 \\
  $ncOS$ $cut$ & -1.22$\; \pm \;$0.06 & -2.00$\; \pm \;$0.06 & 0.40 \\
\hline
\hline
&&$\overline{M_R}$&\\
\hline
$ncOS$ & -1.79$\; \pm \;$0.03 & -2.54$\; \pm \;$0.03 & 0.25 \\
  $ncOS$ $cut$ & -1.68$\; \pm \;$0.04 & -2.65$\; \pm \;$0.03 & 0.24 \\
 \hline
\hline 
&&$\overline{M_J}$&\\
\hline
\hline
$ncOS$ & -2.24$\; \pm \;$0.02 & -2.96$\; \pm \;$0.02 & 0.14 \\
 $ncOS$ $cut$ & -2.18$\; \pm \;$0.02 & -3.03$\; \pm \;$0.02 & 0.14 \\
   \hline
\end{tabular}
\end{center}
\end{table}
}

\onltab{16}{
\begin{table}[ht]
\caption{Theoretical quadratic period-luminosity relations for fundamental
noncanonical pulsators.}\label{table:noncan-quad-PL} 
\begin{center}
\begin{tabular}{lllll}
 \hline
\hline
Case & $a$ & $b$ & $c$ & $\sigma$\\  
\hline 
&&$\overline{M_B}$&\\
\hline
$ncOS$ & -0.87$\; \pm \;$0.20 & -2.78$\; \pm \;$0.36 & 0.41$\; \pm \;$0.15 & 0.41 \\
  $ncOS$ $cut$ & -2.45$\; \pm \;$0.29 & 0.42$\; \pm \;$0.56 & -1.13$\; \pm \;$0.26 & 0.39 \\
   \hline
\hline 
&&$\overline{M_V}$&\\
\hline
$ncOS$ & -1.23$\; \pm \;$0.15 & -3.01$\; \pm \;$0.26 & 0.29$\; \pm \;$0.11 & 0.29 \\
  $ncOS$ $cut$ & -2.39$\; \pm \;$0.21 & -0.67$\; \pm \;$0.4 & -0.84$\; \pm \;$0.19 & 0.28 \\
\hline
\hline 
&&$\overline{M_R}$&\\
  \hline
$ncOS$ & -1.47$\; \pm \;$0.12 & -3.12$\; \pm \;$0.22 & 0.25$\; \pm \;$0.09 & 0.25 \\
  $ncOS$ $cut$ & -2.44$\; \pm \;$0.17 & -1.14$\; \pm \;$0.33 & -0.7$\; \pm \;$0.16 & 0.24 \\
   \hline
\hline
&&$\overline{M_I}$&\\
  \hline
$ncOS$ & -1.72$\; \pm \;$0.10 & -3.22$\; \pm \;$0.18 & 0.22$\; \pm \;$0.08 & 0.21 \\
 $ncOS$ $cut$ & -2.54$\; \pm \;$0.15 & -1.56$\; \pm \;$0.28 & -0.58$\; \pm \;$0.13 & 0.2 \\
\hline
\hline
&&$\overline{M_J}$&\\
  \hline
$ncOS$ & -2.07$\; \pm \;$0.07 & -3.28$\; \pm \;$0.13 & 0.14$\; \pm \;$0.05 & 0.14 \\
  $ncOS$ $cut$ & -2.64$\; \pm \;$0.10 & -2.13$\; \pm \;$0.20 & -0.42$\; \pm \;$0.09 & 0.14 \\
\hline
\hline
&&$\overline{M_K}$&\\
  \hline
$ncOS$ & -2.35$\; \pm \;$0.04 & -3.36$\; \pm \;$0.07 & 0.06$\; \pm \;$0.03 & 0.08 \\
  $ncOS$ $cut$ & -2.66$\; \pm \;$0.06 & -2.72$\; \pm \;$0.11 & -0.25$\; \pm \;$0.05 & 0.08 \\
   \hline
\end{tabular}
\end{center}
\end{table}
}

\begin{figure}
\bigskip
\centering
\includegraphics[width=8cm]{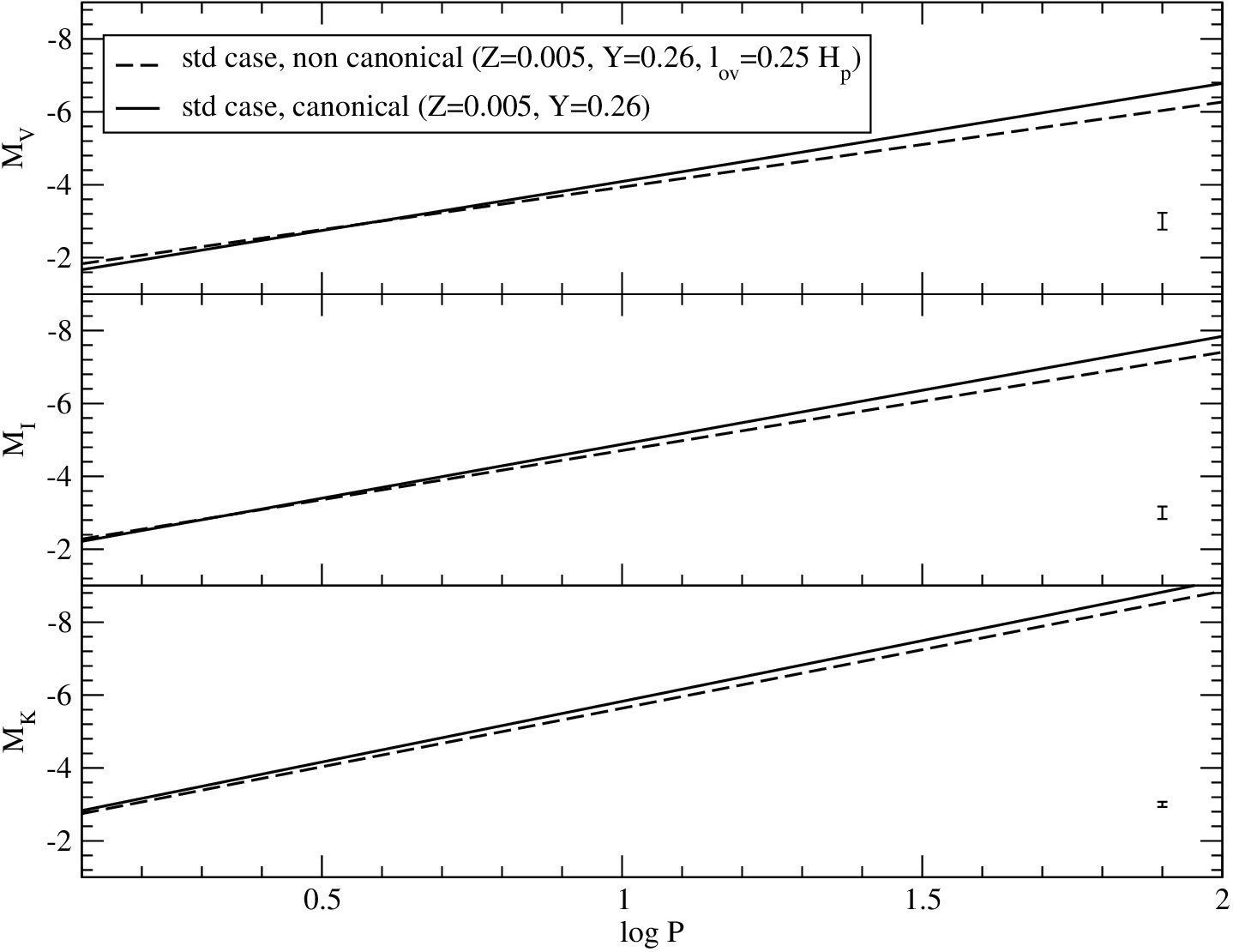}
\caption{Comparison, in the V, I and K photometric bands, 
  between the present canonical and noncanonical linear PL relations  (solid and dashed lines).}
\label{fig:PL-VIK-confronto-can-e-OS}
\end{figure}

A comparison between the canonical (case {\em std} of
Table~\ref{table:relazioniPL}) and the noncanonical $PL$ linear relation in
the V, I, K bands (see Table~\ref{table:relazioniPLlineari-OS}) is shown in
Fig.~\ref{fig:PL-VIK-confronto-can-e-OS}, where the errorbars refer to the intrinsic dispersions of the canonical relations. We notice that both the zero point
and the slope of the $PL$ relations are affected by the inclusion of mild
overshooting. The noncanonical V band relation predicts brighter
magnitudes than the canonical ones in the very short period range and fainter
magnitudes than the canonical ones for periods longer than about 5 days,
whereas the noncanonical I and K band $PL$ predict almost identical magnitudes to the canonical ones at the shortest periods and fainter magnitudes than their
canonical counterparts elsewhere, with the effect increasing
toward longer periods. At a period of 10 days, the effect of applying the
noncanonical relations instead of the canonical ones would be of about 
0.15, 0.17 and 0.19
mag in the V, I and K band respectively, whereas at 30 days it
would be as large as about 0.32, 0.29 and 0.24 in the V, I and K band
respectively.

\begin{table}
\bigskip
\caption{Theoretical $PL$ relations for fundamental noncanonical pulsators (see text).
}      
\label{table:relazioniPLlineari-OS} 
\centering 
\begin{tabular}{l l l l}
  \hline\hline
Case & $a$ & $b$ & $\sigma$ \\
  \hline
         &      &$\overline{M_V}$&\\
\hline
 $ncOS$ & -1.60$\; \pm \;$0.04 & -2.34$\; \pm \;$0.03 & 0.29 \\
  $ncOS$ $cut$ & -1.47$\; \pm \;$0.04 & -2.47$\; \pm \;$0.04 & 0.28 \\
\hline
             &  &$\overline{M_I}$&\\
\hline
$ncOS$ & -2.01$\; \pm \;$0.03 & -2.7$\; \pm \;$0.02 & 0.21 \\
  $ncOS$ $cut$ & -1.92$\; \pm \;$0.03 & -2.8$\; \pm \;$0.03 & 0.20 \\
\hline
             &  &$\overline{M_K}$&\\
\hline
$ncOS$ & -2.43$\; \pm \;$0.01 & -3.21$\; \pm \;$0.01 & 0.08 \\
  $ncOS$ $cut$ & -2.39$\; \pm \;$0.01 & -3.25$\; \pm \;$0.01 & 0.08 \\
\hline
\end{tabular}
\end{table}

Referring to the upper panel of Fig.~\ref{fig:os025}, we report 
in Table~\ref{table:osh-m-e-p-minmax}
the minimum and maximum masses with blue loops that cross the red edge of the instability strip, together with the corresponding values of the minimum and maximum period.

\begin{table}
\caption{M$_{min}$, M$_{max}$, P$_{min}$, P$_{max}$ (as in Table~\ref{table:mminmmax}, \ref{table:mminmmax}) for $l_{\rm ov}=0.25 H_{\rm p}$}     
\label{table:osh-m-e-p-minmax}   
\centering                     
\begin{tabular}{l l l l l }       
\hline\hline
Case & M$_{min}$  & M$_{max}$ & P$_{min}$  & P$_{max}$ \\
&(M$_\odot$)&(M$_\odot$)&(days)&(days)\\
\hline  
XII ($l_{\rm ov}=0.25 H_{\rm p}$)  &4  &9 &4.1 & 72.4  \\
\hline                         
\end{tabular}
\end{table}

\subsection{Mass loss}\label{sec:massloss}

\begin{figure}
\bigskip
\centering
\includegraphics[width=6.5cm]{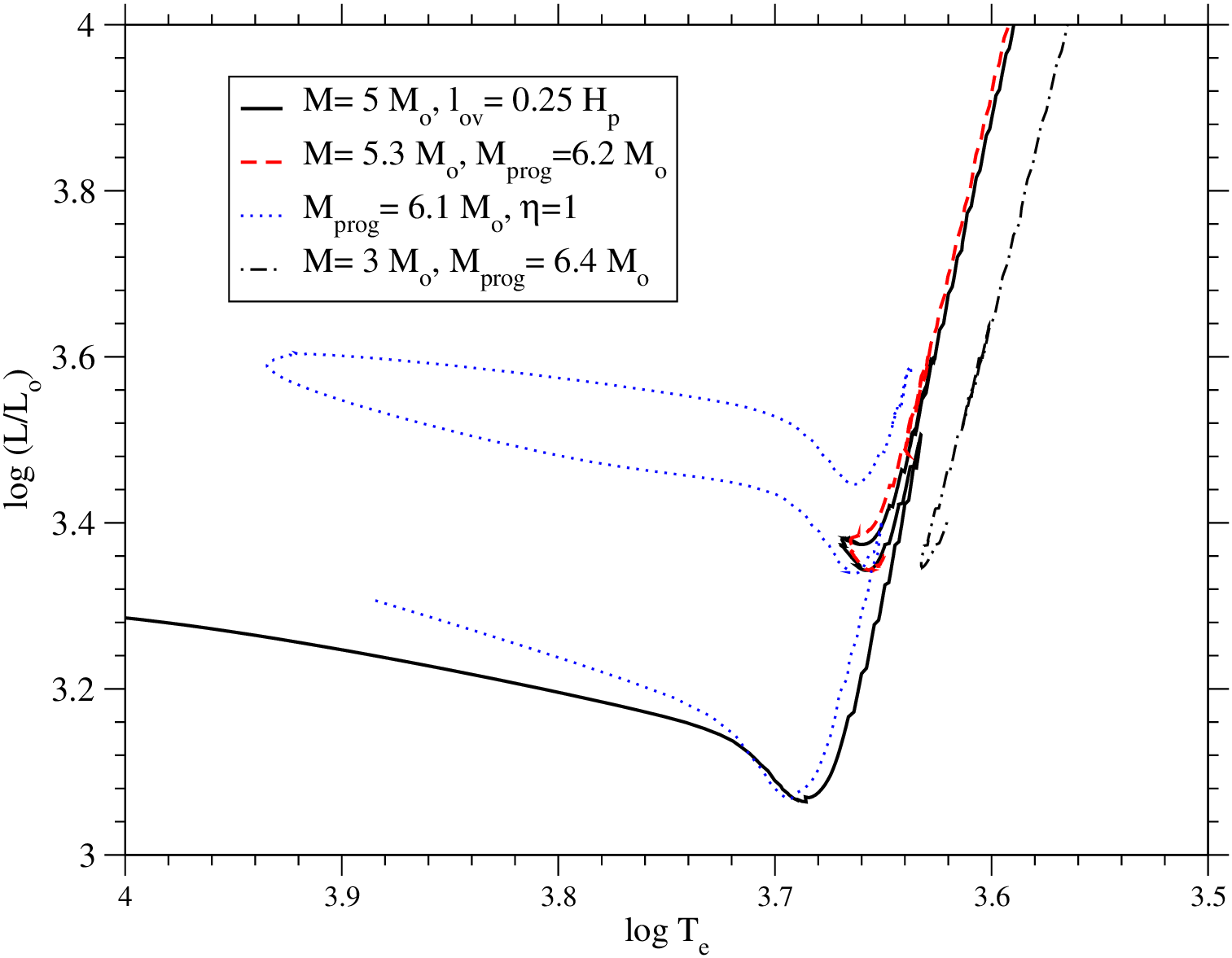}
\caption{Comparison of the minimum luminosity during the central He burning phase
for a 5 M$_{\odot}$ model with $l_{\rm ov}=0.25 H_{\rm_p}$ and models of different
original masses with different massloss prescriptions.}
\label{fig:massloss}
\end{figure}

The possibility that mass loss might solve the long-standing 
problem of the mass discrepancy of Cepheids between the value 
inferred from pulsational and evolutionary models has been 
suggested by various authors \citep[see e.g.][]{bcm02,c05}. 
The argument is that in order 
to solve the problem - that is, to provide higher luminosity for He-burning 
stars of a given mass - a larger He-core mass is needed. 
As previously discussed, this is precisely the effect of the 
convective core-overshooting. On the other hand, a similar 
result can be mimicked by a standard model affected by significant
 mass loss.
In order to check this, we computed the evolution
of a 5 M$_{\odot}$ star taking into account the
effect of a mild core overshooting ($l_{\rm ov}=0.25 H_{\rm p}$) during the
H-burning phase. We attempted to reproduce the minimum luminosity (L$^{ov}_{He}$)
of this model during the central He-burning by standard models affected by mass loss.
The mass loss process is not yet fully understood and 
it is still lacking satisfactory knowledge of its efficiency 
along the various evolutionary phases.
 As a consequence, we are forced to adopt different mass loss prescriptions 
in the following numerical experiments in order to study 
the problem. 
Thus we computed the evolution of stars with mass in the range between
6.0 and 7 M$_{\odot}$ with different mass loss prescriptions 
adopting the classical Schwarzschild criterion
to determine the convective core boundary; the results, discussed below,
are shown in Fig.~\ref{fig:massloss}.
 The models were evolved with constant mass during the main sequence and subgiant phase, while
a substantial mass loss, with a constant rate, was imposed during the
red giant phase. Once the total mass of the models reached the
desired value, the subsequent evolution was performed at constant mass.
Following such a procedure, we computed for example the evolution of
the central He-burning phase of a 5.3 M$_{\odot}$ with a
progenitor star of 6.2 M$_{\odot}$, whose behavior in the HR-diagram
is like that of our target, the 5 M$_{\odot}$ with core overshooting.
However, further numerical experiments showed that the result
is not unique; models with different initial masses and
different prescriptions for the mass loss have a minimum luminosity
during the central He-burning very similar to L$^{ov}_{He}$. 
Such a result was expected since the Cepheid $ML$ relation depends 
 appreciably on both the mass-loss rate and the evolutionary 
phase at which it is turned on. 
An example is given by a star with initial mass of 6.1 M$_{\odot}$ evolved
from the ZAMS, with the classical Reimers mass loss rate (with $\eta$=1);
even if this model describes a much
 more extended loop in color in the HR-diagram, the minimum luminosity of the central He
burning phase is the same. The target luminosity can also be reproduced
by a more extreme model: a 3 M$_{\odot}$ with a progenitor of
6.4 M$_{\odot}$. In this case, the color of the He-burning model is
significantly redder than the model with core overshooting.
These few numerical examples prove that it does not make sense to try to
compute a $ML$ relationship for standard models with mass loss, unless
a priori mass loss prescriptions have been fixed.
Moreover, the mass-loss solution of the Cepheid mass discrepancy 
sounds rather ad hoc, at least until a physical process responsible 
for the huge required mass loss is found and understood. 

\section{Conclusion}\label{sec:concl}

On the basis of an updated set of evolutionary and pulsation models
for Classical Cepheids in the LMC, we have investigated the effect
of the uncertainties on the chemical composition and on the physical
assumptions adopted in the codes
on the most relevant pulsation observables and thus on the theoretical calibration of the
 Cepheid distance scale.
We found that present uncertainties on relevant nuclear reaction rates
have only a negligible effect on evolutionary and pulsational theoretical prediction. 
Moreover the uncertainties in the metal and helium abundances
affect the results of the
evolutionary computations but do not significantly changethe
pulsation scenario. On the other hand, the  still present uncertainties on the
efficiency of the overshooting phenomenon in the previous H burning
phase and on the mass-loss rates are found to be the most important source of
uncertainty in the theoretical Cepheid mass-luminosity and period-luminosity
relations.  
In particular, using a theoretical $PL$ relation that relies on the
assumption of mildly overshooting evolutionary models, one infers distances
that are significantly shorter than the values obtained when a canonical
theoretical $PL$ is used, especially in the optical bands.  
Therefore we conclude that the uncertainty on the Cepheid $ML$
relation is expected to significantly affect the Cepheid calibration
of the extragalactic distance scale and in turn the evaluation of the
Hubble constant. The application of the theoretical analysis performed in this 
paper to LMC Cepheid data will be addressed in a forthcoming paper (Valle et al. 2009 in preparation).

\begin{acknowledgements}
We thank our anonymous referee for his/her valuable comments that improved the quality of our paper. We also warmly thank S. Shore for a careful reading of the manuscript. Financial support for this study was provided by PRIN-INAF 2006 (PI G. Clementini). 
\end{acknowledgements}

\bibliographystyle{aa}

\end{document}